\documentclass[aps,prb,twocolumn,letterpaper,nobibnotes,superscriptaddress,longbibliography]{revtex4-1}
\usepackage{graphicx}
\usepackage{dcolumn}
\usepackage{bm}
\usepackage{amsmath}
\usepackage{amssymb}
\usepackage{diagbox}
\usepackage{subfigure}
\usepackage{color}
\usepackage{cancel}
\usepackage{soul}
\usepackage{ulem}

\begin{document}

\title{Bosonic fractional quantum Hall conductance in shaken honeycomb optical lattices without flat bands}
\author{Shiwan Miao}
\author{Zhongchi Zhang}
\author{Yajuan Zhao}
\author{Zihan Zhao}
\author{Huaichuan Wang}
\affiliation{Department of Physics and State Key Laboratory of Low Dimensional Quantum Physics, Tsinghua University, Beijing, 100084, China}
\author{Jiazhong Hu}
\email{hujiazhong01@ultracold.cn}
\affiliation{Department of Physics and State Key Laboratory of Low Dimensional Quantum Physics, Tsinghua University, Beijing, 100084, China}
\affiliation{Frontier Science Center for Quantum Information, Beijing, 100084, China}

\date{\today}

\begin{abstract}
We propose a scheme to realize bosonic fractional quantum Hall conductance in shaken honeycomb optical lattices. 
This scheme does not require a very flat band, 
and the necessary long-range interaction relies on the s-wave scattering which is common in many ultracold-atom experiments.
By filling the lattice at 1/4 with identical bosons under Feshbach resonance, two degenerate many-body ground states share one Chern number of 1 and exactly correspond to the fractional quantum Hall conductance of 1/2. 
Meanwhile, we prove that the fractional quantum Hall state can be prepared by adiabatically turning on the lattice shaking, and the fractional conductance is robust in the shaken lattice. This provides an easy way to initialize and prepare the fractional quantum Hall states in ultracold-atom platforms, 
and it paves a new way to investigate and simulate strong-correlated quantum matters with degenerate quantum gas.
\end{abstract}

\maketitle

\section{Introduction}
The fractional quantum Hall (FQH) effect is one of the most fascinating phenomena in past decades \cite{Stormer1999RMP}, where multiple many-body ground states share one integer Chern number and effectively each of the band obtains a fractional number to characterize the conductivity.
In previous studies \cite{Stormer1999RMP,CooperBook,Wen2017RMP}, it was proved that the FQH effect can be achieved either in fermions or bosons. However, due to the fermionic nature of electrons in conventional materials, the FQH effect has only been observed experimentally in fermions. 
Thus, this leaves an open question about how to realize the bosonic FQH effect experimentally, for example, to realize and probe the Hall conductance corresponding to one half.

With the development of quantum simulations \cite{Bloch2008RMP,Cooper2019RMP}, the platforms of ultracold atoms now provide good opportunities to study and simulate strong-correlated many-body systems \cite{Lukin2005PRL,hafezi2007fractional,Cooper2013PRL,2011Fractional,neupert2011fractional,PhysRevA.100.053624,Sun2011PRL,Tang2011PRL,Grushin2014PRL,PhysRevLett.110.185301}, particularly for bosonic FQH effect.
There are many pioneer experiments in realizing non-trivial Chern numbers or large synthetic gauge fields \cite{miyake2013realizing,Aidelsburger2013,Jotzu2014,aidelsburger2015measuring,Monika2011PRL,Monika2020realization,beeler2013spin,stuhl2015visualizing,lin2009synthetic} in order to reach regimes of strong correlations. 
Inspired by previous studies of FQH effect in ``Haldane-like'' models \cite{2011Fractional,neupert2011fractional,haldane1988model,Jotzu2014}, we find that the bosonic FQH conductance can be achieved and experimentally initialized in shaken honeycomb optical lattices without synthetic magnetic fields. 
This scheme relies on the Feshbach resonance at s-wave scatterings \cite{Chin2010RMP} which does not require special long-range interactions.

Meanwhile, compared to previous studies \cite{2011Fractional,neupert2011fractional,Sun2011PRL,Tang2011PRL} requiring a flat band where the band gap is much larger than the band width, our scheme is realized with the nearest-neighbor-hopping model and does not require a very flat band. Our band gap is almost the same as the band width. 
This avoids special designs for high-order hoppings for far-apart lattice sites to flatten the bands, and reduces the complexity of Hamiltonian engineering. 
Furthermore, we find that the states with FQH conductance can be prepared by adiabatically turning on the shaking of static optical lattices which is topologically trivial. This simplifies the procedures to initialize the FQH states in optical lattices. We believe this scheme will inspire new opportunities to experimentally study the bosonic FQH effect in an easier way.

\section{Shaken honeycomb optical lattices and single-body topology}

\begin{figure*}[htbp]
\centering
\includegraphics[width=\textwidth]{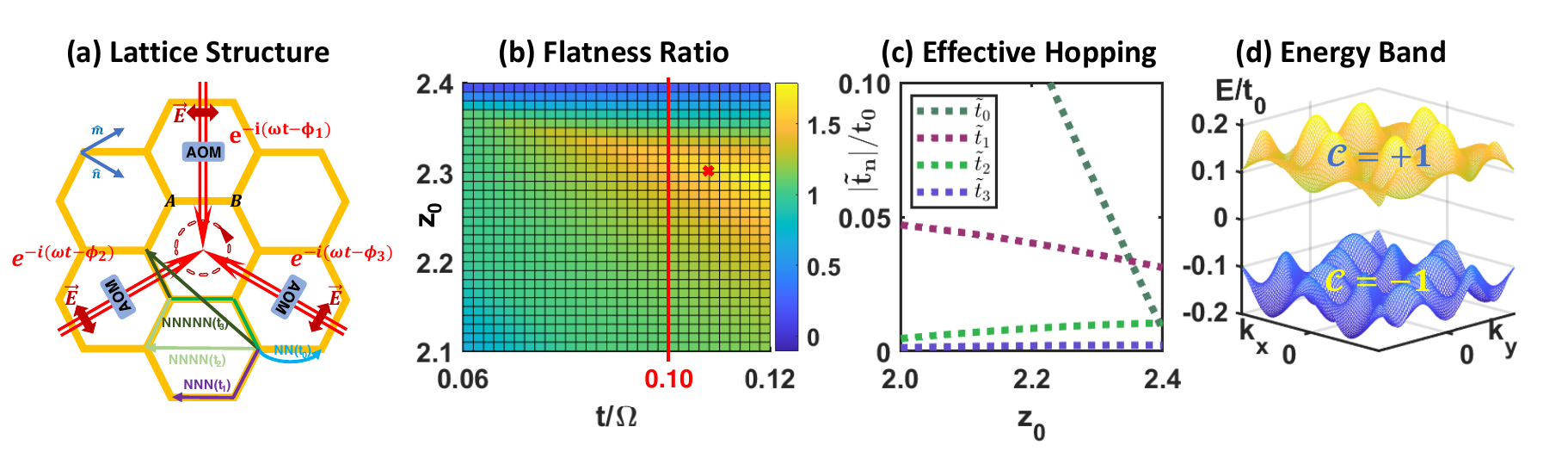}
\caption{\label{fig:single}
(a) The schematic of shaken honeycomb optical lattices. The definition of the $m$- ($n$-) axis of lattice coordinate and the hopping terms are marked on the lattice.
When the phases of three beams are modulated by $\phi_1=\phi_A \cos(\Omega t+\pi/2)$, $\phi_2=\phi_A \cos(\Omega t-\pi/6)$, and $\phi_3=\phi_A \cos(\Omega t+5\pi/6)$, the lattice sites move along a counterclockwise circular trajectory whose angle frequency is $\Omega$ and orbital radius is proportional to $\phi_A$.
(b) The band flatness ratio versus modulation parameters $z_0$ and $t_0/\Omega$. (c) Effective tunnelings $\tilde t_m$ versus $z_0$ at $t_0/\Omega=0.1$ (red line in panel b).  (d) The single-particle band at $z_0=2.3$, $t_0/\Omega=0.108$ (red cross in panel b), where each band has a non-zero Chern number. }
\end{figure*}

The honeycomb optical lattice is formed by the interference of three red-detuned lasers at the same frequency, which have a relative angle at 120 degrees  and  are in one incident plane (Fig.~\ref{fig:single}a). The polarizations of lattice beams are parallel to the incident plane. The dipole potential $V_{op}$ of the optical lattice can be written in the form of
\begin{eqnarray}
\label{eqn:Lattice}
V_{op}&=&-{V_D}\left|e^{-iky}\hat{x}+e^{ik({\sqrt{3}\over 2}x+{1\over 2}y)}(-{1\over 2}\hat{x}+{\sqrt{3}\over 2}\hat{y}) \right. \nonumber \\
& &\left. +e^{ik({-\sqrt{3}\over 2}x+{1\over 2}y)}(-{1\over 2}\hat{x}-{\sqrt{3}\over 2}\hat{y})\right|^2  \nonumber \\
&=&-V_D\left[3-\cos\sqrt{3}kx-\cos\sqrt{3}k(x/2+\sqrt{3}y/2)\right. \nonumber \\
& &\left.-\cos\sqrt{3}k(-x/2+\sqrt{3}y/2)\right],
\end{eqnarray}
where $V_D$ is the trap depth in our definition, $k=2\pi/\lambda$ is the wave vector, and $\lambda$ is the wavelength of lattice lasers. This gives a lattice spacing $a=\lambda/2\sqrt 3$, corresponding to a hexagon whose side length is $\lambda/2\sqrt 3$. 
When cold atoms are trapped in honeycomb lattices, they can be described by a tight-binding model, whose Hamiltonian $H_0$ is 
$H_0/\hbar=-\sum_{\langle i, j\rangle} t_0 \hat{c}^\dagger_i \hat{c}_j$ where $t_0$ is the nearest-neighbor hopping amplitude; $\hat{c}_i^{\dagger}$ ($\hat{c}_i$) is the creation (annihilation) operator on lattice cite $i$; $\langle\cdot\rangle$ corresponds to the summation of all nearest-neighbor sites.

In order to create non-trivial topological bands in tight-binding honeycomb lattice, we apply a periodic modulation on the phases of the lattice beams  to break the time reversal symmetry, where phases of each laser beam are modulated according to $\phi_1=\phi_A \cos(\Omega t+\pi/2)$, $\phi_2=\phi_A \cos(\Omega t-\pi/6)$, and $\phi_3=\phi_A \cos(\Omega t+5\pi/6)$. 
This creates a shaken lattice of which each site $i$ follows a circular motion $\vec{r}_{i}(t)$ where $\vec{r}_{i}(t)=\vec{r}_{i,0}-A[\cos(\Omega t)\hat{x}+\sin(\Omega t)\hat{y}]$ and $A$ is orbital radius of the circular trajectory. The lattice shaking alternates the original Hamiltonian $H_0$ into a time-dependent form $\hat{H}'(t)$, and
\begin{equation}
\hat{H}'(t)/\hbar= -\sum_{\langle i,j\rangle}e^{iz_{ij}\sin(\Omega t+{\phi}_{ij})}t_0\hat{c}^{\dagger}_{i}\hat{c}_{j}+h.c.,
\label{Hamiltonianshaken}
\end{equation}
where $\langle i,j\rangle$ corresponds to a pair of nearest-neighbor lattice sites, $z_{ij}=m_{a}\Omega A\rho_{ij}/\hbar$, $m_{a}$ is the mass of an atom, and $\rho_{ij}e^{i\phi_{ij}}=(\vec{r}_{i,0}-\vec{r}_{j,0})\cdot(\hat{x}+\hat{y}e^{-i\pi/2})$.
Here $z_{ij}$ is the ratio of $m_{a}\Omega^2 A\rho_{ij}$ to $\hbar \Omega$, where the numerator is the product of the centrifugal shaking force $m_{a}\Omega^2 A$ and the distance $\rho_{ij}$, and the denominator is the Floquet energy $\hbar \Omega$.

A periodic Hamiltonian is decomposed into a Fourier transformation that $\hat{H}'(t) = \sum_{l \in \mathbb{Z}} \hat{H}_l e^{il \Omega t}\label{eq:Fourier}$.
When $\Omega$ is large, we obtain an effective time-indepedent Floquet Hamiltonian $\hat{H}_{fl}$ based on high frequency expansion method \cite{goldman2014periodically,Eckardt2015NJP,2003Effective,2015Universal}. 
We rewrite the effective Floquet Hamiltonian $\hat{H}_{fl}$ based on the nearest-neighbor (NN), next-nearest-neighbor (NNN), next-next-nearest-neighbor (NNNN), and next-next-next-nearest-neighbor (NNNNN) hoppings, \textit{i.e.}
\begin{eqnarray}
\hat{H}_{fl}/\hbar&=-&\sum_{\langle i,j\rangle}\tilde{t}_{0}\hat{c}^\dagger_i \hat{c}_j-\sum_{\langle i,j \rangle_{2}}\tilde{t}_{1}\hat{c}^\dagger_i \hat{c}_j-\sum_{\langle i,j \rangle_{3}}\tilde{t}_{2}\hat{c}^\dagger_i \hat{c}_j \nonumber \\
& &-\sum_{\langle i,j \rangle_{4}}\tilde{t}_{3}\hat{c}^\dagger_i \hat{c}_j +h.c.,
\end{eqnarray}
where $\langle\cdot\rangle$, $\langle \cdot \rangle_{2}$, $\langle \cdot \rangle_{3}$, and $\langle \cdot \rangle_{4}$ correspond to the summations of the NN, NNN, NNNN, and NNNNN sites. 
We plot $\tilde{t}_{m}$ in Fig.~\ref{fig:single}c and the detailed formula of $\tilde{t}_{m}$ can be found in the appendix. The effective NNN hopping amplitude has an imaginary part, which breaks the time-reversal symmetry, opens the band gap, and gives a non-zero Chern number to each band.
The band flatness ratio is usually used to characterize the potential of many-body topology \cite{Fractional2013CRP}, where the ratio is defined by the band gap divided by the band width of the ground band.
In Fig.~\ref{fig:single}, we present the flatness ratio, hopping strength and energy bands under different parameters.
Usually a flat band is required to show the dominated bosonic FQH effect \cite{2011Fractional,neupert2011fractional}, while  
flattening the band is a challenge in ultracold-atom experiments since the intrinsic long-range hoppings are strongly suppressed for remote lattice sites.
In our scheme, the flatness ratio is less than 2 where the gap is near the same as the band width. 
We find it is still suitable to realize the FQH conductance of 1/2.

\section{Many-body topology with strongly interacting bosons}

Now we switch our description from single-body physics to many-body physics. To achieve the fractional conductance, we need both strong on-site interactions and finite nearest-neighbor interactions. 
This can be realized by conventional s-wave Feshbach resonance \cite{Chin2010RMP}.
For a given trap depth $V_D$ in Eq. \ref{eqn:Lattice}, we calculate the Wannier functions of each site by the methods in Ref.~\cite{2011Maximally} (See Appendix \ref{App:Wannier} for more details). The on-site interaction $U$ is obtained by the integral
$
U={4\pi\hbar^2 a_s\over m_{a}}\int d^{3}\bm{r} w^{\dagger}(\bm{r}) w^{\dagger}(\bm{r})w(\bm{r})w(\bm{r}) 
$
where $a_s$ is the scattering length and $m_a$ is the mass of one atom. $w(\bm{r}-\bm{r}_i)$ is the Wannier function centered at the position $\bm{r}_i$. Usually in Bose-Hubbard models, we only keep the on-site interactions and ignore the NN interaction $V_1$. However, once the scattering length $a_s$ approaches to infinity under Feshbach resonance, the NN interaction $V_1$ becomes significant with a form of \cite{zhai_2021}
\begin{eqnarray}
V_{1}&=&{8\pi\hbar^2 a_s\over m_{a}}\int d^3 r w^{\dagger}(\bm{r}) w^{\dagger}(\bm{r}-\bm{a})w(\bm{r}-\bm{a})w(\bm{r}) \nonumber \\
&=&2U{\int d^3 r w^{\dagger}(\bm{r}) w^{\dagger}(\bm{r}-\bm{a})w(\bm{r}-\bm{a})w(\bm{r})\over\int d^{3}r w^{\dagger}(\bm{r}) w^{\dagger}(\bm{r})w(\bm{r})w(\bm{r})}.
\end{eqnarray}
Here $\bm{a}$ corresponds to the relative vector between two nearest-neighbor sites.

In rubidium 85, there is a Feshbach resonance at 155.3~G \cite{Courteille1998PRL, Claussen2003PRA} with a width around 11~G and a background scattering length -441$a_0$ where $a_0$ is the Bohr radius. Considering that magnetic-field fluctuations can be controlled within 1~mG in most of labs, the scattering length can be tuned up to 10000$a_0$ with a relative uncertainty less than 0.25\%. This offers the opportunity that the on-site interaction reaches the hard-core regime while the nearest-neighbor interaction is still important in the tight-binding model. 
In Table~\ref{tab:Wannier}, we list the hopping coefficients and interaction strength versus $V_D$ in unshaken lattices. 
At the condition of $a_s=10000a_0$ and $V_D=28E_r$ ($E_r = {\hbar^2 k^2 \over 2m_{a}}$), the original $t_0$ is $0.031E_r$ and the on-site interaction $U$ is $88E_r$, which satisfies $U\gg t_0$. 
$U$ strongly repulses any doublons in one site. Meanwhile, the NN interaction $V_1$ is $0.56\hbar t_0$ and is significant compared with the hopping amplitudes.

\begin{table}[h]
\caption{\label{tab:Wannier}
$V_1$ versus the trap depth $V_D$.
We calculate the intrinsic NN tunneling $t_0$, NNN tunneling $t_1$, and NNNN tunneling $t_2$ of the original unshaken lattice, where  $t_1$ and $t_2$ are usually negligible compared with $t_0$. 
Here we assume there is a perpendicular trap tightly confining cold atoms into 2D degenerate gas, and the tight trap is described by a harmonic trap with a vibrational frequency at 50~kHz. 
}
\begin{ruledtabular}
\begin{tabular}{ccccccc}
$V_D/E_r$&$\hbar t_{0}/E_r$&$\hbar t_{1}/t_{0}$&$10^4 t_{2}/t_{0}$&$U/E_r$&$V_1/2E_r$ &$V_1/t_0$\\
\colrule
20&0.059&-0.020&$16$&71.34&0.034&1.16\\
22&0.050&-0.016&$11$&75.47&0.024&0.96\\
24&0.043&-0.014&$7.4$&79.41&0.017&0.78\\
26&0.037&-0.011&$5.2$&83.19&0.012&0.64\\
28&0.031&-0.0096&$3.6$&86.82&0.0088&0.58\\
30&0.027&-0.0081&$2.6$&90.31&0.0063&0.48\\
32&0.023 &-0.0068&$1.8$&93.68&0.0047&0.40\\
\end{tabular}
\end{ruledtabular}
\end{table}

In the strong-correlation regime, the single-particle band description cannot capture the actual physics. 
Therefore, using the twisted boundary condition, we write out the many-body Floquet Hamiltonian $\hat H_{fl-many}$ of hard-core bosons with the nearest-neighbor interaction $V_1$ under the trap depth $V_D=28E_r$. Here the twisted boundary condition is $\psi(\bm{r_i}+L_m\hat{\bm{m}})=e^{i\theta_x}\psi(\bm{r_i})$ and $\psi(\bm{r_i}+L_n \hat{\bm{n}})=e^{i\theta_y}\psi(\bm{r_i})$ where $L_m$($L_n$) is the lattice size along $m$- ($n$-) axis and the axes are marked in Fig.~\ref{fig:single}a. At $\theta_{x}=\theta_{y}=0$, the twisted boundary condition is the same as the periodic boundary condition.  $\hat H_{fl-many}$ is a giant-sized sparse matrix (see Appendix \ref{App:Hamiltonian} for more information).
Then we apply the exact diagonalization to calculate the lowest four energy levels and their corresponding many-body bands in the case of 6 bosons in 24 lattice sites ($\hat H_{fl-many}$ has a size around $10^5\times 10^5$), where the bands are characterized by $\theta_x$ and $\theta_y$ instead of the quasi-momenta $k_x$ and $k_y$.

By scanning the Floquet parameter $z_0$ and NN interaction $V_1$, there are some regions that the lowest two many-body bands cross each other while they are away from the third band. 
In Fig.~\ref{fig:manybody}a, we present the energy difference between the second and third energy states ($E_3$-$E_2$) under the twisted boundary condition. Here we take the inherent long-range hopping beyond the tight-binding models ($t_1$ and $t_2$) into accounts in calculating the 0th-order effective Hamiltonian, and neglect them in higher-order expansions. 
We mark the candidates for the FQH conductance at 1/2 in Fig. \ref{fig:manybody}a according to the gap opening. In Fig. \ref{fig:manybody}c, we plot the many-body bands versus $\theta_x$ and $\theta_y$ in this phase regime. It shows that two lowest bands cross each other and are away from the third band.

To further verify the conductance, we calculate the total Chern number $\mathcal{C}$ of the lowest two bands. Here $\mathcal{C}$ equals
${1\over 2\pi}\sum_{K=1,2}\int d\theta_x d\theta_y F_{xy,K}(\theta_x,\theta_y)$ where $F_{xy,K}(\theta_x,\theta_y)=\textrm{Im}(\langle \frac{\partial\psi_{K}}{\partial\theta_y}|\frac{\partial\psi_{K}}{\partial\theta_x}\rangle-\langle \frac{\partial\psi_{K}}{\partial\theta_x}|\frac{\partial\psi_{K}}{\partial\theta_y}\rangle)$ is the Berry curvature of the $K$-th state in the ground state manifold. 
We divide the whole $\theta_x$-$\theta_y$ space into 20$\times$20 pieces, and calculate the energies and the discrete Berry curvatures \cite{Fukui2005Chern} in Fig.~\ref{fig:manybody}c and d. The two lowest bands share one integer Chern number $\mathcal{C}=-1$ together, corresponding to FQH conductance of 1/2. We calculate cases in lattices of different sizes to show robustness of our scheme against the finite size effect (Fig. \ref{fig:manybody}b), while the size of $\hat{H}_{fl-many}$ increases dramatically ($\hat{H}_{fl-many}$ has a size of $10^8\times10^8$ for 36 sites).

\begin{figure}[tb]
\centering
\includegraphics[width=0.46\textwidth]{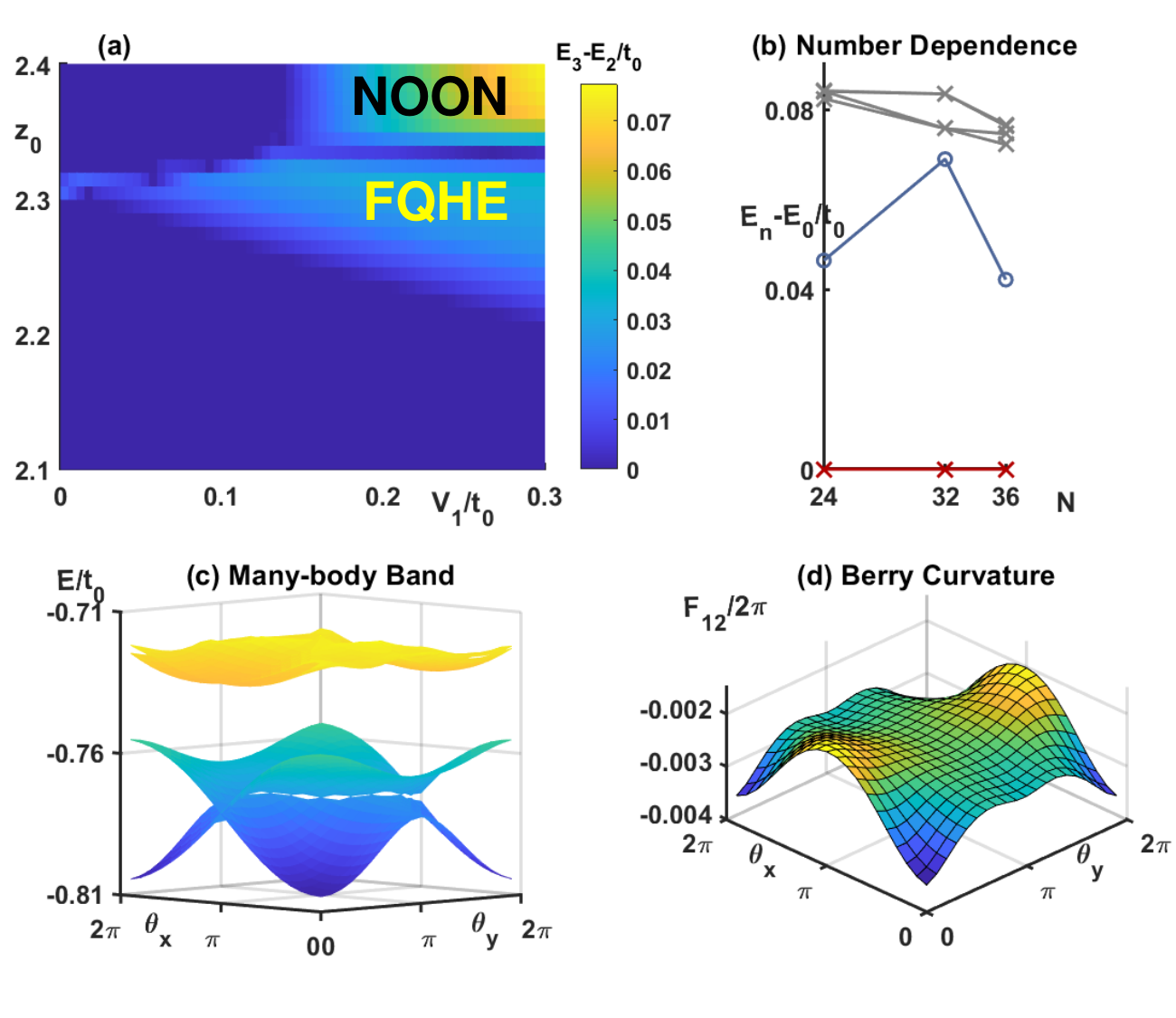}
\caption{\label{fig:manybody}(a) Different phases distinguished by many body gap $E_3-E_2$. We also label out NOON phase here. While the system is in NOON states, the honeycomb lattice is decomposed into two individual sets of triangle lattices. The atoms fill one of the triangle lattices and leaves the other empty. The actual ground states are the superpositions of these two possibilities, like a NOON state. (b) The variation of many-body energy levels with lattice sites $N$ under $V_D=28E_{r}$. Here it shows the same behavior as Ref \cite{2011Fractional} and maintains the FQH band gap. (c) Many-body bands at $V_D=28E_r$. (d) Discrete Berry curvature of two lowest states while the integral is -1.}\label{fig:manybody}
\end{figure}

\section{Robustness against the higher band excitation and Floquet heating}

The tight-binding Hamiltonian in Eq.~\ref{Hamiltonianshaken} only considers the contributions from s-bands, while a large scattering length may cause the s-band to mix with higher-excited bands. 
We follow the same treatment in Ref \cite{Suppressing2021Esslinger} which successfully describes the interband transitions in optical lattices. We build a model with two sites, two particles, and both s-bands and p-bands to estimate the contributions from higher bands. The detailed numerical calculations and analyses are listed in the appendix \ref{App:TwoModel}, and only the conclusion is presented in the main text. The dominant interband transitions induced by a large scattering length is that two neighboring particles in s-bands will scatter to p-bands together due to collisions. The coupling matrix element of this process is less than 0.1$E_r$, but the energy detuning is more than 10$E_r$. This suggests that there will be a suppressing factor more than $10^{-4}$ for higher-band mixture, which is negligible in our systems.

Besides the phenomena of higher-band mixture due to a large scattering length,
the Floquet modulation may also cause the transitions to the excited bands. 
A particle at site-$i$ may absorb $\Delta n_{p}$ Floquet "photons", and be excited to a higher band. The effective 
coupling strength $Q_{\Delta n_{p}}$ to excite a particle to higher bands via a resonant $\Delta n_{p}$-photon process is (see Appendix \ref{App:Heating} for detailed calculations)
\begin{equation}
Q_{\Delta n_{p}}=B_{\Delta n_{p}}(\frac{z_{0}}{z_{0,th}})^{\Delta n_{p}-1}.
\end{equation}
Here $B_{\Delta n_{p}}$ is a coefficient whose magnitude has weak dependence on $\Delta n_{p}$ and we list the detailed form in Appendix \ref{App:Heating}.  The threshold $z_{0,th}$ is a dimensionless quantity which is approximately the photon number $\Delta n_p$ divided by the Euler's number $e$. In the case of $V_D=28E_r$, it requires at least 28 photons to excite a particle to the higher bands so $z_{0,th}$ is around 10, while $z_0$ in our proposal is 2.3. Then $(\frac{z_{0}}{z_{0,th}})^{\Delta n_{p}-1} $ in the equation provides a factor of $10^{-19}$. Therefore, the interband heating caused by Floquet modulation is negligible in our scenario.

\section{Adiabatic preparation of FQH states}

Another potential question about our model is whether it's appropriate to derive effective Hamiltonian by the high-frequency-expansion method. 
To eliminate this concern and prove that we can prepare such FQH states adiabatically in cold-atom platforms, in the following calculations we apply the original time-dependent Hamiltonian, which contains intrinsic long-range hoppings and does not include Floquet treatments.
The original time-dependent Hamiltonian is in the form of
\begin{eqnarray}
H(t)= -&\Big[&\sum_{<i,j>}e^{iz_{ij}\sin(\Omega t+{\phi}_{ij})}t_0\hat{c}^{\dagger}_{i}\hat{c}_{j}\nonumber\\
&+&\sum_{<i,j>_{2}}e^{i z_{ij}\sin(\Omega t+{\phi}_{ij})}t_1\hat{c}^{\dagger}_{i}\hat{c}_{j}\nonumber\\
&+&\sum_{<i,j>_{3}}e^{iz_{ij}\sin(\Omega t+{\phi}_{ij})}t_2\hat{c}^{\dagger}_{i}\hat{c}_{j} \nonumber\\
&+&\sum_{<i,j>_{4}}e^{iz_{ij}\sin(\Omega t+{\phi}_{ij})}t_3\hat{c}^{\dagger}_{i}\hat{c}_{j} +...\nonumber\\
&+&h.c. \Big] +\sum_{\langle i,j\rangle}V_{i,j}\hat{n}_{i}\hat{n}_{j} .
\label{eqn:realH}
\end{eqnarray} 
Here $t_0$, $t_1$, $t_2$ and $t_3$ are intrinsic NN, NNN, NNNN, and NNNNN hopping coefficients for static honeycomb optical lattices (Table I). 
The on-site interaction $U$ does not appear in the expression since we limit the Hilbert space into the states where each site can be filled with one particle at most.
$z_{ij}$ and $\phi_{ij}$ have the same meaning as those in Eq.~\ref{Hamiltonianshaken} and have different values for different hoppings.

Therefore, 
we simulate how to prepare the target FQH state, which is the ground state of the effective Hamiltonian under the twisted boundary condition, with FQH conductance of 1/2 .
Initially, the lattice is static and unshaken, and we start with the ground state $|\psi(z_0=0)\rangle$ under twisted boundary conditions. Then we fix the shaking frequency $\Omega$ at 0.108 and linearly ramp the shaking amplitude $A$ to increase $z_0$ from 0 to 2.3 in
$C$ Floquet-modulation cycles, and we calculate the fidelity, which is the module square of inner products, between the time-evolved state and the target FQH state. 

\begin{figure}[tb]
\centering
\includegraphics[width=0.5\textwidth]{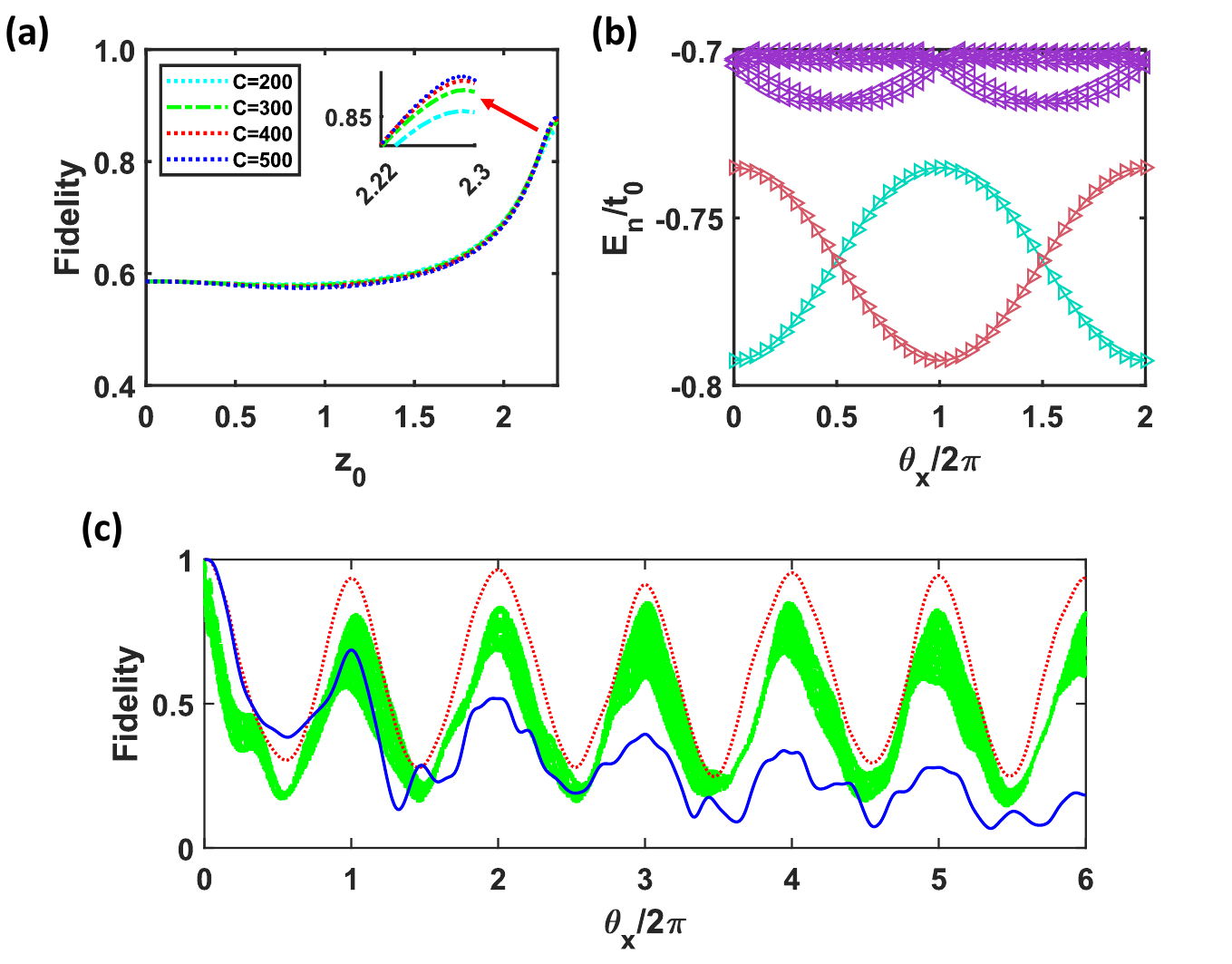}
\caption{(a) Fidelity versus $z_0$ and $C$.
The inset is a zoom-in plot to show the difference near $z_0=2.3$. 
Slowly turning on the Floquet modulation helps the fidelity, and 400 modulation cycles will be slow enough to reach the adiabatic limit.
 (b) The spectrum of energy $E_n$ versus $\theta_x$ for energy-lowest eight states of the effective Hamiltonian.
 The energy is normalized by $t_0$ and $\theta_x$ is changed from 0 to $4\pi$. Two lowest states share the Chern number 1 together, and it is a typical signature of fractional quantum Hall conductance at $1/2$.
(c) Fidelity within the manifold of the lowest two states while changing $\theta_x$. Green belt for the adiabatic-evolved state; red dotted line for the ground state of the effective Hamiltonian; blue solid line for the ground state of a topologically trivial Hamiltonian.}
\label{fig:Prepare}
\end{figure}

In Fig.~\ref{fig:Prepare}a, we plot the fidelity versus $z_{0}$ and different ramping rates (or total modulation cycles $C$).
The fidelity at the end of ramping reaches over 85\% for both cycles $C=300$ and $400$, and approaches to a constant while $C>400$. 
We find if we ignore $\tilde t_3$ (NNNNN hoppings) in the effective Hamiltonian, the fidelity drops below 80\%. Although $\tilde t_3$ is only 1/30 of other major terms, it hurts the calculations of fidelities. It suggests that the ground state of the effective Hamiltonian is not identical to the experimentally-prepared state by adiabatic ramping, and this causes the fidelity to fall below 100\%.

Following the discussion of FQH conductance in Ref. \cite{niu1985quantized}, the ground state and the first excited state cross each other in the many-body energy spectrum while they are isolated from the higher excited states. Besides this level isolation, one of the 1/2-FQH states evolves into the other one when the boundary condition $\theta_x$ is changed  by $2\pi$.
 Then, $4\pi$ instead of 2$\pi$ becomes the period for the state to evolve back to itself. 
In Fig.~\ref{fig:Prepare}b, we show that the states of the effective Hamiltonian have a $4\pi$-period and the level isolation.

Therefore, for the adiabatic-evolved states without tight-binding approximations, we apply the same arguments to prove the FQH conductance.
We use State I to denote the adiabatic-evolved state during the preparation.
We tune $\theta_x$ by $2\pi$ very slowly, and State I evolves and cannot come back to itself since the gap is closed during the change of boundary conditions.
We extract the orthogonal component at $2\pi$ change and refer it as State II. This state maintains a good fidelity (always above 75\%) with the first excited state of the effective Hamiltonian.

We change $\theta_x$ for multiple $2\pi$ (Fig.~\ref{fig:Prepare}c) and calculate the overall fidelity within this manifold of State I and II for the actual scenario. In Fig.~\ref{fig:Prepare}c, we show how the fidelity changes while the boundary condition is being varied (green belts). When $\theta_x$ leaves 0 or $2\pi$, the overlap between State I and State II decreases since the state is evolving. When $\theta_x$ reaches integer times of $2\pi$, the fidelity comes back again. After a few $2\pi$, the fidelity is stably oscillating and does not decay while $\theta_x$ is varied.
The shape of belts is due to rapid Floquet modulations. Within each modulation, the fidelity is oscillating and the long-time-scale fidelity behaves as an area instead of a line. Here the fidelity does not exactly come back to 1 and we believe this is mainly due to the numerical errors of the giant matrix and data storage. 

To convince this point, we numerically calculate another two cases for direct comparisons. 
One case is for the effective Hamiltonian under Floquet theory (red dotted line), and another is for a topological-trivial Hamiltonian (blue solid line). For the effective Hamiltonian, the change of fidelity behaves very similar to the case of adiabatic-evolved state. It is robust and does not decay while we continue changing $\theta_x$. 
Since we analytically solve this model, the ideal case should be that the fidelity reaches 1 again when $\theta_x$ is integer times of $2\pi$.
The fidelity below 1 is due to the issues of numerical errors of a large-size matrix. The same reason applies to the case of the adiabatic-evolved state. Therefore, the robustness of the fidelity in the manifold of State I and State II proves the properties of FQH conductance.
For the case of the topological-trivial Hamiltonian, we find the fidelity is decreasing while we change $\theta_x$. It indicates that the wave function is leaking out from this manifold. There is no gap separating the energy-lowest two states with the other bands, which cannot support FQH conductance.

\section{Conclusions}

In conclusion, we find a scheme realizing the bosonic FQH conductance of 1/2 in optical lattices without flat band. By circularly shaking the optical lattice and applying Feshbach resonance, the system displays a fractional quantum Hall conductance of 1/2. We show that the state with this conductance can be experimentally prepared by adiabatic turning on the lattice shaking. It provides a convenient and robust way to investigate the bosonic FQH effect \cite{PhysRevB.85.235137, PhysRevB.41.12838, PhysRevLett.114.186801, PhysRevLett.122.166801, PhysRevA.85.033620, PhysRevA.98.063621} in cold-atom experiments.

We acknowledge financial supports from the National Natural Science Foundation of China under grant no. 11974202, and 92165203.




\begin{appendix}
\section{Effective Floquet Hamiltonian}\label{App:Effective}
In this section, we derive the effective Floquet Hamiltonian for shaken optical lattices. The lattice moves along a trajectory $\bm{r}_l(t)$ where the subscript $l$ corresponds to the displacement of the whole lattice. In the lattice coordinate, there is an inertial force $\bm{F}(t)=-m \bm{\ddot{r}}_l(t)$ applying to each atom whose mass is $m$, and this leads to a site-dependent potential $-\bm{F}(t)\cdot \bm{r}_i$. 
Therefore, the Hamiltonian in the lattice coordinate has a form of
\begin{eqnarray}
\hat{H}(t)&=&-\sum_{ i,j} \hbar t_{ij} \hat{c}^{\dagger}_i \hat{c}_j-\sum_{i}(\bm{F}(t)\cdot \bm{r}_i)\hat{c}^{\dagger}_i \hat{c}_i \nonumber\\
&+&\sum_{{\langle}i,j\rangle }V_{i,j}\hat{n}_{i}\hat{n}_{j} + {1\over 2}\sum_{i}U_{i}\hat{n}_{i}(\hat{n}_{i}-1)
\label{Eq:SHamiltonian}
\end{eqnarray}
where $i$ or $j$ may represent any lattice site while $\langle i,j \rangle$ corresponds to a nearest-neighbor pair of lattice sites. By introducing a unitary transformation 
\begin{equation}
\hat{U}(t)=\exp[{i \over \hbar}\sum_{i}(-m_{a} \bm{\dot{r}}_l(t) \cdot \bm{r}_{i}) \hat{c}_{i}^{\dagger} \hat{c}_i],
\end{equation}
the Hamiltonian is converted into
\begin{equation}
\hat{H}^{\prime}(t)=\hat{U}^{\dagger}(t) \hat{H}(t) \hat{U}(t) -i \hbar \ \hat{U}^{\dagger}(t)\partial_{t} \hat{U}(t).
\label{Eq:SUnitary}
\end{equation}
The first term on the right hand side leaves the inertial-force potential and interaction terms unchanged, and the second term cancels out the inertial-force potential.   
By applying the Baker-Compell-Hausdorff formula, the hopping terms after the transformation become  
\begin{equation}
\hat{U}^{\dagger}(t) (-\sum_{i,j } \hbar t_{ij} \hat{c}^{\dagger}_i \hat{c}_j) \hat{U}(t)=-
\hbar \sum_{i,j } t_{ij}\ \hat{c}^{\dagger}_i \hat{c}_{j} e^{{i\over\hbar}m_{a} \bm{\dot{r}}_{l}(t)\cdot\bm{r}_{ij}},
\end{equation}
where $\bm{r}_{ij}=\bm{r}_i-\bm{r}_j$ is the relative position between the site $i$ and $j$.
And the Hamiltonian after the transformation is 
\begin{eqnarray}
\hat{H}^{\prime}(t)&=&-\hbar \sum_{ i,j } t_{ij}\ \hat{c}^{\dagger}_i \hat{c}_{j} e^{{i\over\hbar}m_{a} \bm{\dot{r}}_{l}(t)\cdot\bm{r}_{ij}}\nonumber\\
&+& \sum_{\langle i,j \rangle}V_{i,j}\hat{n}_{i}\hat{n}_{j} + {1\over 2}\sum_{i}U_{i}\hat{n}_{i}(\hat{n}_{i}-1).
\end{eqnarray}
 When the lattice moves along a circular trajectory with a radius $A$, \textit{i.e.} $\bm{r}_l(t)=\bm{r}_{l,0}-A[\cos (\Omega t)\bm{\hat{x}}+\sin(\Omega t)\bm{\hat{y}}]$. 
Then we introduce the symbols $\rho_{ij}$ and $\phi_{ij}$ to simplify the equation, whose definitions are $\rho_{ij}e^{i\phi_{ij}}=(\vec{r}_{i,0}-\vec{r}_{j,0})\cdot(\hat{x}+\hat{y}e^{-i\pi/2})$. Then $\hat{H}^{\prime}(t)$  is 
\begin{eqnarray}
\hat{H}^{\prime}(t) &=&-\hbar \sum_{i,j } t_{ij}\ \hat{c}^{\dagger}_i \hat{c}_{j} e^{{i\over\hbar}m_{a} A\Omega \rho_{ij}\sin(\Omega t+\phi_{ij})}\nonumber\\
&+&\sum_{\langle i , j \rangle}V_{i,j}\hat{n}_{i}\hat{n}_{j} + {1\over 2}\sum_{i}U_{i}\hat{n}_{i}(\hat{n}_{i}-1).
\end{eqnarray}
Then we use a dimensionless parameter $z_{ij}=m_{a}A\Omega \rho_{ij}/\hbar$ to characterize the system. In the main text $z_0$ is the nearest-neighbor $z_{ij}$ and the higher order of $z_{ij}$ can be derived based on this formula and the value of $z_0$.
A periodic Hamiltonian can be Fourier-decomposed by the Jacobi-Anger method into $\hat{H}'(t) = \sum_{n \in \mathbb{Z}} \hat{H}_n e^{in \Omega t}$, and the $n$-th order Fourier term for $\hat{H}^{\prime}(t)$ is
\begin{eqnarray}
\hat{H}_{n}&=&-\hbar\sum_{i,j }e^{in\Omega t} J_{n}(z_{ij}) e^{in\phi_{ij}}t_{ij}\ \hat{c}^{\dagger}_i \hat{c}_{j} \nonumber\\
&+& \delta_{n0}(\sum_{\langle i,j \rangle}V_{i,j}\hat{n}_{i}\hat{n}_{j} + {1\over 2}\sum_{i}U_{i}\hat{n}_{i}(\hat{n}_{i}-1))\textcolor{red}{,}
\end{eqnarray}
where $J_n(x)$ is the $n$-th Bessel function.

In the high frequency region, the time-dependent Hamiltonian can be approximated by a time-independent Floquet Hamiltonian $\hat{H}_{fl}$ based on high frequency expansion method \cite{goldman2014periodically,Eckardt2015NJP,2003Effective,2015Universal} with a form of
$\hat{H}_{fl}=\hat{H}_{0\Omega }+\hat{H}_{1\Omega}+\hat{H}_{2\Omega}+\ldots$.
Here $\hat{H}_{0\Omega }$, $\hat{H}_{1\Omega }$, and $\hat{H}_{2\Omega }$ are obtained via the commutation relations of $\hat{H}_n$, \textit{i.e.}
\begin{eqnarray}
\hat{H}_{0\Omega}&&= \hat{H}_0,  \\
\hat{H}_{1\Omega}&&= \frac{1}{\hbar\Omega} \sum_{n=1}^{\infty} \frac{1}{n} [\hat{H}_n,\hat{H}_{-n}],    \\
\hat{H}_{2\Omega}&&= \frac{1}{2\hbar^2\Omega^2} \sum_{n=1}^{\infty} \frac{1}{n^2}([\hat{H}_n,\hat{H}_0],\hat{H}_{-n}]+ h.c.)\nonumber\\
&&+\frac{1}{3\hbar^2\Omega^2} \sum_{n,n'=1}^{\infty} \frac{1}{nn'}([\hat{H}_n,[\hat{H}_{n'},\hat{H}_{-n-n'}]]\nonumber\\
&&-[\hat{H}_n,[\hat{H}_{-n'},\hat{H}_{-n+n'}]]+ h.c.). \label{eqn:HFE2}
\end{eqnarray}

Considering an unshaken lattice with only nearest-neighbor (NN) hopping $t_0$, the commutator of two NN hopping produces a new hopping term with longer range, \textit{i.e.} $[\hat{c}^{\dagger}_{i}\hat{c}_{j},\hat{c}^{\dagger}_{j}\hat{c}_{k}]=\hat{c}^{\dagger}_{i}\hat{c}_{k}$. Therefore, the Floquet Hamiltonian has nearest-neighbor (NN), next-nearest-neighbor (NNN), next-next-nearest-neighbor (NNNN), and next-next-next-nearest-neighbor (NNNNN) hoppings, with a form of.
\begin{eqnarray}
\hat{H}_{fl}/\hbar&=-&\sum_{\langle i,j\rangle}\tilde{t}_{0}\hat{c}^\dagger_i \hat{c}_j-\sum_{\langle i,j \rangle_{2}}\tilde{t}_{1}\hat{c}^\dagger_i \hat{c}_j-\sum_{\langle i,j \rangle_{3}}\tilde{t}_{2}\hat{c}^\dagger_i \hat{c}_j \nonumber \\
& &-\sum_{\langle i,j\rangle_{4}}\tilde{t}_{3}\hat{c}^\dagger_i \hat{c}_j +h.c.,
\label{eqn:HamiltonianEff}
\end{eqnarray}
where $\langle\cdot\rangle$, $\langle\cdot\rangle_{2}$, $\langle\cdot\rangle_{3}$, and $\langle\cdot\rangle_{4}$ correspond to the summations of the NN, NNN, NNNN, and NNNNN sites. The effective hopping amplitudes are 
\begin{widetext}
\begin{eqnarray}
\tilde t_{0}&=&t_{0}J_{0}(z_{0})+\frac{2t^3_0}{\Omega^2}\sum_{s=1}^{\infty}\frac{1}{s^2} J_{s}(z_{0})J_{0}(z_{0})J_{-s}(z_{0})\times[2\cos(2s\pi/3)-2\cos(s\pi/3)\nonumber\\
& &+2\cos(2s\pi/3) -3\cos(s\pi)+1]+\frac{4t^{3}_0}{3\Omega^{2}}\sum_{s,s'=1}^{\infty}\frac{1}{ss'}\{J_{s}(z_{0})J_{s'}(z_{0})J_{-s-s'}(z_{0})\times 
\nonumber\\
& &[2\cos(2s\pi/3-s'\pi/3)-2\cos(s\pi+s'\pi/3)+2\cos(-2s\pi/3-s'\pi)
\nonumber\\
& &-2\cos(s\pi/3+s'\pi)+\cos(s'\pi)-\cos(s\pi+s'\pi)]-(s'\rightarrow -s')\},
\\
\tilde t_{1}&=&-\frac{2it^{2}_0}{\Omega}\sum_{s=1}^{\infty}\frac{1}{s}J_{s}(z_{0})J_{-s}(z_{0})\sin(s\pi/3),
\\
\tilde t_{2}&=&\frac{4t_0^3}{\Omega^2}\sum_{s=1}^{\infty}\frac{1}{s^2} J_{s}(z_{0})J_{0}(z_{0})J_{-s}(z_{0})\times[\cos(2s\pi/3)-\cos(s\pi/3)]\nonumber \\
& &+\frac{8t_0^{3}}{3\Omega^{2}}\sum_{s,s'=1}^{\infty}\frac{1}{ss'}\{J_{s}(z_{0})J_{s'}(z_{0})J_{-s-s'}(z_{0})\times[\cos(2s\pi/3+s'\pi/3) \nonumber \\
& &-\cos(s\pi/3-s'\pi/3)]-(s'\rightarrow -s')\},
\\
\tilde t_{3}&=&\frac{2t_0^3}{\Omega^2}\sum_{s=1}^{\infty}\frac{1}{s^2} J_{s}(z_{0})J_{0}(z_{0})J_{-s}(z_{0})\times[1-\cos(s\pi/3)]\nonumber \\
& &+\frac{4t_0^{3}}{3\Omega^{2}}\sum_{s,s'=1}^{\infty}\frac{1}{ss'}\{J_{s}(z_{0})J_{s'}(z_{0})J_{-s-s'}(z_{0})\times[\cos(-s'\pi/3) \nonumber \\
& &-\cos(-s\pi/3-s'\pi/3)]-(s'\rightarrow -s')\}.
\label{eqn:tunneling}
\end{eqnarray}
\end{widetext}
Here we have not taken the interaction terms into accounts. The influence of the NN interaction $V_1$ on the Floquet Hamiltonian is proportional to ${1\over\Omega^2}$ and produces number-dependent NNN hopping. 
In the high frequency region, The NNN hopping induced by $V_1$ is less than one tenth of that in Eq.~13 so we neglect it in the calculation of effective hopping.
This approximation affects the calculation of the ground state of effective Hamiltonian, so we simulate the wavefunction under exact time-dependent Hamiltonian in the main text and show that the negligence is reasonable.  
Because we are interested in the scenario of hard-core bosons ($U\rightarrow \infty$), it requires a high energy cost for two particels to occupy the same site and the Floquet photons cannot provide such a large energy. In the region where the modulation frequency is much smaller than the energy scale of interaction, the high frequency expansion method is not applicable. In Appendix D, we show that it is safe to confine the Hilbert space in the subspace composed by single-occupation states despite the Floquet modulation. Therefore, the on-site interaction does not appear in the Hamiltonian.

\section{Wannier functions in honeycomb optical lattices}\label{App:Wannier}

We calculate the ground-state Wannier function in honeycomb optical lattices in order to characterize the on-site and nearest-neighbor interactions. 
First, we calculate the Bloch functions $|\psi_{1,\bm{k}}\rangle$ and $|\psi_{2,\bm{k}}\rangle$ of the first and the second energy bands at a $50\times50$ $\bm{k}$-point mesh. The Wannier function is the Fourier transform of Bloch function with a form of 
\begin{equation}
|\bm{R}n\rangle = {V \over (2\pi)^3}\int_{\textrm{FBZ}} d\bm{k}|\psi_{n,\bm{k}}\rangle e^{-i\bm{k}\cdot\bm{R}}
\end{equation}
where $\bm{R}$ is lattice vector, $|\bm{R}n\rangle$ is the $n$-th Wannier function in  the primitive cell at $\bm{R}$, the integral domain is the first Brillouin zone (FBZ). 

The choice of Bloch function has a gauge freedom that permits the replacement of $|\psi_{n,\bm{k}}\rangle$ by $e^{i\theta(\bm{k})}|\psi_{n,\bm{k}}\rangle$. Therefore, a good gauge should make the derivative of Bloch function $\nabla|\psi_{n,\bm{k}}\rangle$ well defined in the FBZ. If there are $J$ bands having crossover with each other, the gauge freedom is generalized to a unitary transform $U_{mn}^{(\bm{k})}$ among $J$ Bloch functions 
\begin{equation}
|\tilde\psi_{n\bm{k}}\rangle = \sum_{m=1}^{J} |\psi_{m\bm{k}}\rangle U_{mn}^{(\bm{k})}.
\end{equation}

Here, the gauge is determined by the projection method \cite{Marzari1997PRB,2011Maximally}. We use the ground state of the harmonic trap localized at $A$- or $B$- site of honeycomb cells as the trial functions $|h_1\rangle$ and $|h_2\rangle$. We define a matrix $A(\bm{k})_{mn}=\langle \psi_{m,\bm{k}}|h_n\rangle$, and the unitary transform $U_{mn}^{(\bm{k})}$ can be represented by the singular value decomposition of $A(\bm{k})=U_s(\bm{k})D_s(\bm{k})V^{\dagger}_s 
\bm{k})$, where $U_s(\bm{k})$ and $V_s(\bm{k})$ are unitary and $D_s(\bm{k})$ is diagonal. Then, the gauge matrix $U_{mn}^{(\bm{k})} $ is $U_{mn}^{(\bm{k})} = U_s(\bm{k})V_s^{\dagger}(\bm{k})$.

Then, we use the results above to calculate the required inputs of Wannier90 \cite{Pizzi2020} and optimize the output Wannier functions. 
We find that the spread of optimized Wannier functions reaches the minimal which verifies that the Wannier function obtained from the projection method is the maximally localized Wannier function (MLWF). It's required that the MLWF should be real all over the space, and the maximal magnitude of imaginary parts in our result is $10^{-13}$ less than the maximal values of real parts. It supports that the non-vanishing imaginary parts are caused by numerical precision and do not hurt our calculations.

\begin{figure}
\centering
\includegraphics[width=0.5\textwidth]{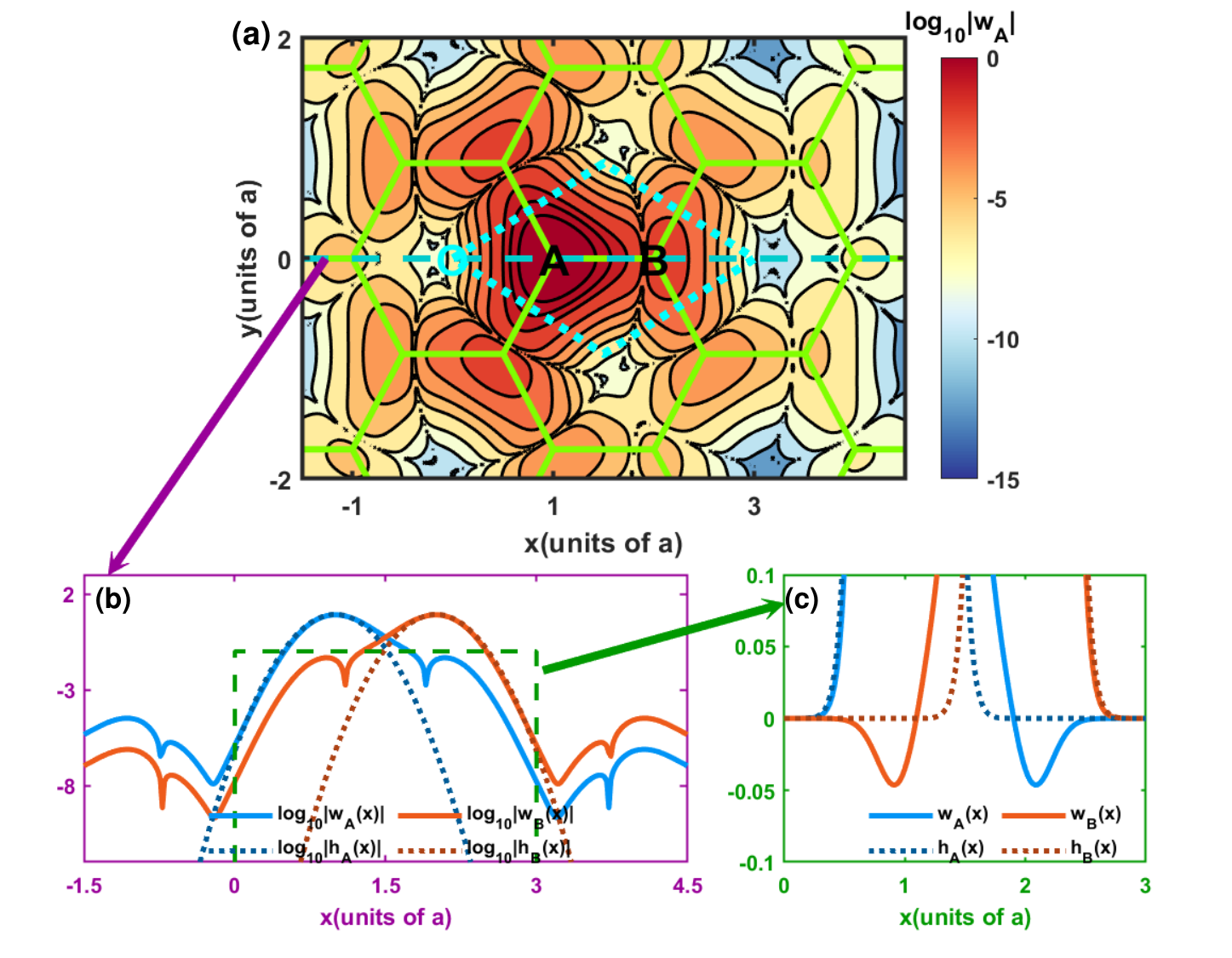}
\caption{(a) MLWF centered at A site. The trap depth $V_D=30E_r$. Here $a$ is the lattice constant corresponding to the distance between the nearest $A$ site and $B$ site. (b) Comparison of MLWFs and trial functions (Gaussian functions from harmonic traps). (c) A zoom-in plot of panel b for comparison of MLWFs and trial functions around 0. All the panels are plotted in the logarithmic scale for the wave function values.}
\label{fig:MLWF}
\end{figure}

In Fig.~\ref{fig:MLWF}(a), the MLWF centered at $A$ site $w_{A}(r)$ is plotted in a logarithmic scale. In Fig.~\ref{fig:MLWF}(b) and (c), the MLWFs along $x$-axis are plotted, and they are compared with the trial functions $h_{A}(r)$ and $h_{B}(r)$. Both MLWFs and trial functions are normalized so that the squared-integral in the $x-y$ plane is 1.

The calculation of interactions in BH model requires three-dimensional Wannier function, so we assume a harmonic trap along $z$-axis with a vibrational frequency $\omega_{z} = 2\pi \times 50$ kHz. We use the ground state wave function $w_A$(z) of the harmonic trap along $z$-axis for the third dimension. 
The integral contribution along $z$-axis is
$\int dz w_{A}^{\dagger}(z) w_{A}^{\dagger}(z)w_{A}(z)w_{A}(z)=\sqrt{m_{a}\omega_{z}\over \hbar}$, where $A$ corresponds to $A$ site. We use rubidium 85 as an example and the integral contribution is $4.16 \ \mu \textrm{m}^{-1}$. In Table~\ref{tab:overlap}, we present the integral in the $x-y$ plane and the corresponding interaction at the scattering length $a_s=10000a_0$ ($a_0$ is the Bohr radius). The integral for the on-site overlap is 
\begin{equation}
I_{1}=\int dxdy w_{A}^{\dagger}(x,y) w_{A}^{\dagger}(x,y)w_{A}(x,y)w_{A}(x,y).
\end{equation}
The integral of the NN overlap is 
\begin{equation}
I_2=\int dz w_{A}^{\dagger}(x,y) w_{B}^{\dagger}(x,y)w_{B}(x,y)w_{A}(x,y).
\end{equation}
The hopping terms are numerically acquired by fitting the Bloch bands to the tight-binding model with higher hopping terms \cite{Tight2013PRA}. The results are listed in the main text.

\begin{table}
\caption{\label{tab:overlap}
The overlap contributions in the $x$-$y$ plane for on-site and nearest-neighbor versus the trap depth $V_D$. The integral of overlap $I_1$ and $I_2$ is normalized to the units of $\lambda^{-2}$, where $\lambda$ is the wavelength of lasers for optical lattices. The energy is normalized to the unit of the recoil energy $E_r=h^2/(2m\lambda^2)=\frac{\hbar^2 k^{2}}{2m_{a}}$.
}

\begin{ruledtabular}
\begin{tabular}{ccccc}
$V_D$ $(E_r)$&$I_1$ $(\lambda^{-2})$&$I_2$ $(\lambda^{-2})$&$U$ $(E_r)$&$V_1{/2}$ $(E_r)$\\
\colrule
20&25.9009&0.0123&71.34&0.0339\\
22&27.3995&0.0087&75.47&0.0240\\
24&28.8305&0.0062&79.41&0.0171\\
26&30.2013&0.0044&83.19&0.0121\\
28&31.5182&0.0032&86.82&0.0088\\
30&32.7867&0.0023&90.31&0.0063\\
32&34.0115&0.0017&93.68&0.0047\\
\end{tabular}
\end{ruledtabular}
\end{table}

\section{Two-particle, Two-site and Two-band Model}
\label{App:TwoModel}

For bosons in lattices, the field operator is 
\begin{equation}
\hat{\psi}(\bm{r})=\sum_{m,\bm{R}_i}\hat{b}_{m,i}w_{m}(\bm{r}-\bm{R}_{i}),
\end{equation}
where $\hat{b}_{m,i}$ is the annihilation operator of bosons with site index $i$ and band index $m$, $w_{m}(\bm{r}-R_{i})$ is the corresponding Wannier function.
Considering a delta-function potential, the general form of the lattice model is \cite{zhai_2021}
\begin{equation}
\hat{H}=-\sum_{ijm}t_{ij}^{m}\hat{b}^{\dagger}_{m,i}\hat{b}_{m,j}
+{1\over 2}\sum_{iji'j'}^{mnm'n'}U_{ijkl}^{mnm'n'}\hat{b}^{\dagger}_{m,i}\hat{b}^{\dagger}_{m',i'}\hat{b}_{n,j}\hat{b}_{n',j'},
\label{eqn:BHmodel}
\end{equation}
where
\begin{equation}
t_{ij}^{m}=-\int d^{3}\bm{r} w_{m}(\bm{r}-\bm{R}_{i})\left( -\frac{\hbar^2 \nabla^2}{2m}+V_{op}(\bm{r})\right)w_{m}(\bm{r}-\bm{R}_j),
\end{equation}
and
\begin{eqnarray}
U_{ijkl}^{mnm'n'}&&={4\pi\hbar^{2} a_{s}\over m_{a}}\int d^{3}\bm{r}\nonumber\\
&&w_{m}(\bm{r}-\bm{R}_{i})w_{n}(\bm{r}-\bm{R}_{j})w_{m'}(\bm{r}-\bm{R}_{k})w_{n'}(\bm{r}-\bm{R}_{l}).\nonumber\\
\label{eqn:interaction}
\end{eqnarray}

\begin{figure}
\centering
\includegraphics[width=0.5\textwidth]{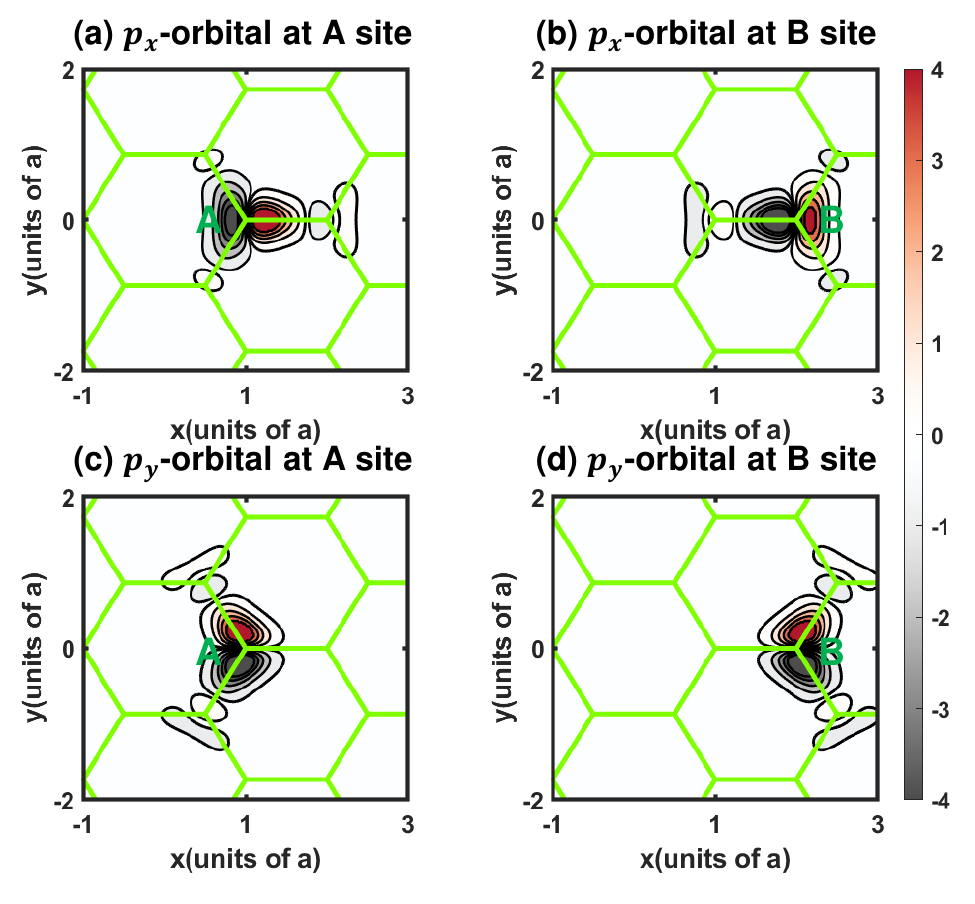}
\caption{$p_{x}$-orbital and $p_{y}$-orbital at $A$- or $B$-sites.}
\label{fig:Pwave}
\end{figure}

Following the previous reference \cite{Suppressing2021Esslinger}, we use the two-particle, two-site, and two-band model to estimate the band mixing due to large scattering length. We calculate the Wannier functions of the p-bands and plot them In Fig.~\ref{fig:Pwave} for both $A$- and $B$-sites. Because $p_y$-orbitals are anti-symmetric in $y$-direction and more localized along $x$-direction, the major contributions of band mixing with s-bands come from the $p_x$-orbital. Therefore, we keep the $s$-band and $p_x$-band in $A$- and $B$-sites to estimate the band mixture. The lattice Hamiltonian is further simplified into
\begin{eqnarray}
H&=&-t^g_{AB}(\hat{b}^\dagger_{g,A} \hat{b}_{g,B}+\hat{b}^\dagger_{g,B} \hat{b}_{g,A})-t^e_{AB}(\hat{b}^\dagger_{e,A} \hat{b}_{e,B}+\hat{b}^\dagger_{e,B} \hat{b}_{e,A}) \nonumber \\
& &-t^g_{AA}(\hat{b}^\dagger_{g,A} \hat{b}_{g,A}+\hat{b}^\dagger_{g,B} \hat{b}_{g,B})-t^e_{AA}(\hat{b}^\dagger_{e,A} \hat{b}_{e,A}+\hat{b}^\dagger_{e,B} \hat{b}_{e,B}) \nonumber \\
& &+{1\over 2}\sum_{iji'j'\in\{A,B\}}^{mnm'n'\in\{e,g\}}U_{ijkl}^{mnm'n'}\hat{b}^{\dagger}_{m,i}\hat{b}^{\dagger}_{m',i'}\hat{b}_{n,j}\hat{b}_{n',j'}
\end{eqnarray}
Here the index $g$ ($e$) corresponds to the $s$- ($p_x$-) band particles, and the index $A$ ($B$) corresponds to the $A$- ($B$-) site. We set $t^g_{AA}=0$ as the zero energy point, and then $t^e_{AA}$ is characterizing the band separation between $s$- and $p$-bands. In Table \ref{tab:numeric}, we list the typical numerical magnitudes in this model. Since we are interested in the effect on the ground state $\hat{b}^\dagger_{g,A}\hat{b}^\dagger_{g,B}|vac\rangle$, where $|vac\rangle$ is the vacuum state, we focus on the transitions from the ground state to the states with higher energy. In Table \ref{tab:a lot of U}, all the possible first-order transitions are listed with the energy costs and the coupling strengths. The ratios between the coupling strengths and the energy cost are less than 1/100, so the excited populations due to Feshbach resonance are less than $10^{-4}$. Therefore, the band mixture due to large scattering length is highly suppressed by the off resonant coupling.

\begin{table}
\caption{\label{tab:numeric}
The typical numerical values in the two-particle, two-site and two-band model at $a_s=10000a_0$ and $V_D=28E_r$. The calculation is based on $s$-orbital Wannier functions and $p_x$-orbital Wannier functions at $A$ and $B$-sites. The Wannier function along $z$-direction is the ground state wave function in a harmonic trap with a vibrational frequency at 50~kHz. Here all the parameters are in the unit of the recoil energy $E_r={\hbar^2 k^2\over 2m_{a}}$.
}
\begin{ruledtabular}
\begin{tabular}{ccc}
 &numeric value ($E_{r}$)& \\
\colrule
$\hbar t_{AA}^{e}\approx-8.271$& $U^{egeg}_{ABAB}=0.6409$& $U^{eggg}_{ABAB}=0.0212$\\
$U_{AAAA}^{gggg}=86.82$&$U_{AAAB}^{gggg}=-0.2828$\\
$U_{AAAA}^{egeg}=34.6818$&$U_{AAAB}^{eggg}=-0.2648$\\
$U_{ABAB}^{eeee}=2.1601$&$U_{ABAB}^{eegg}=-0.0676$\\
$U_{AAAA}^{eeee}=55.0342$ & $U^{eegg}_{AAAB}=0.1704$
\end{tabular}
\end{ruledtabular}
\end{table}

\begin{table}
\caption{\label{tab:a lot of U}
The first-order transition channels.
}
\begin{ruledtabular}
\begin{tabular}{ccc}
transition&energy cost&coupling\\
\colrule
$\hat{b}_{g,A}^{\dagger}\hat{b}_{g,B}^{\dagger}|vac\rangle\rightarrow\hat{b}_{g,A}^{\dagger}\hat{b}_{e,B}^{\dagger}|vac\rangle$&$-t_{AA}^{e}+2U^{egeg}_{ABAB}$&$2U_{ABAB}^{eggg}$\\
(or $\hat{b}_{g,B}^{\dagger}\hat{b}_{e,A}^{\dagger}|vac\rangle$)&\\
\hline
$\hat{b}_{g,A}^{\dagger}\hat{b}_{g,B}^{\dagger}|vac\rangle\rightarrow\hat{b}_{g,A}^{\dagger}\hat{b}_{g,A}^{\dagger}|vac\rangle$&$U^{gggg}_{AAAA}$&$U^{gggg}_{AAAB}$\\
(or $\hat{b}_{g,B}^{\dagger}\hat{b}_{g,B}^{\dagger}|vac\rangle$)&\\
\hline
$\hat{b}_{g,A}^{\dagger}\hat{b}_{g,B}^{\dagger}|vac\rangle\rightarrow\hat{b}_{g,A}^{\dagger}\hat{b}_{e,A}^{\dagger}|vac\rangle$&$-t_{AA}^{e}+2U^{egeg}_{AAAA}$&$2U^{eggg}_{AAAB}$\\
(or $\hat{b}_{g,B}^{\dagger}\hat{b}_{e,B}^{\dagger}|vac\rangle$)&\\
\hline
$\hat{b}_{g,A}^{\dagger}\hat{b}_{g,B}^{\dagger}|vac\rangle\rightarrow\hat{b}_{e,A}^{\dagger}\hat{b}_{e,B}^{\dagger}|vac\rangle$&$-2t_{AA}^{e}+2U^{eeee}_{ABAB}$&$2U^{eegg}_{ABAB}$\\
\hline
$\hat{b}_{g,A}^{\dagger}\hat{b}_{g,B}^{\dagger}|vac\rangle\rightarrow\hat{b}_{e,A}^{\dagger}\hat{b}_{e,A}^{\dagger}|vac\rangle$&$-2t_{AA}^{e}+U^{eeee}_{AAAA}$&$U_{AAAB}^{eegg}$\\
(or $\hat{b}_{e,B}^{\dagger}\hat{b}_{e,B}^{\dagger}|vac\rangle$)&\\
\end{tabular}
\end{ruledtabular}
\end{table}

\section{Floquet Heating}\label{App:Heating}

In this section, we estimate the interband excitations caused by Floquet modulations. Since the energy scale for interband excitations is much larger than the modulation frequency, the high frequency expansion method is not applicable. The quasienergy operator and a set of new bases, combining stationary states and Floquet photons, are applied to solve this problem \cite{Eckardt2015NJP}. Although there are lots of interaction terms in Eq.~\ref{eqn:interaction}, only the on-site interaction has larger energy than a Floquet photon. Therefore, we focus on the on-site interaction and ignore the nearest-neighbor interaction.

Similar to Eq.~\ref{Eq:SHamiltonian}, the complete tight-binding Hamiltonian in the moving reference frame is
\begin{eqnarray}
\hat{H}_{c}(t)&=&-\sum_{i,j,m,m'}(\bm{F}(t)\cdot \bm{r}_{i,j}^{m'm})\hat{c}^{\dagger}_{i,m'} \hat{c}_{j,m}\nonumber\\
&-&\sum_{ i,j, m} \hbar t_{ij}^{m} \hat{c}^{\dagger}_{i,m} \hat{c}_{j,m}+{1\over 2}\sum_{i,m}U_{iiii}^{mmmm}\hat{n}_{i,m}(\hat{n}_{i,m}-1).\nonumber\\ \label{eq:D1}
\end{eqnarray}
Here $m$ is band index, $t_{ii}^{m}$ is the mean energy of the $m$-th band, and $\bm{r}_{i,j}^{m'm}=\int d^{3}\bm{r}w_{m'}(\bm{r}-\bm{R}_{i})\ \bm{r}\ w_{m}(\bm{r}-\bm{R}_{j})$. 
If only the contributions from the s-bands are taken into account, $\bm{r}_{i,j}^{ss}$ will equal to $\delta_{ij} \bm{r}_{i}$ where $\bm{r}_{i}$ is the position of the lattice site $i$ in the moving reference frame. Then we obtain the same results as from Eq.~\ref{Eq:SHamiltonian}. 

For interband transitions, there is relation of $\bm{r}_{i,i}^{mm'}\gg \bm{r}_{i,j(\neq i)}^{mm'}$.
Therefore, by inspecting the first term in the right hand side of Eq. \ref{eq:D1}, the excitation by Floquet photons is mainly from the in situ transitions where the particle is still in the same spatial position but jumps to a higher band. Then we will focus on this contribution and demonstrate that it is not a concern.
For simplicity, the repeated subscripts or superscripts will be contracted, \textit{e.g.} $r^{m,m}_{i,i}\rightarrow r^m_{i}$.
The Hamiltonian is simplified to
\begin{widetext}
\begin{eqnarray}
\hat{H}_{c}(t)&=&-\sum_{ i,j ,m} \hbar t_{ij}^{m} \hat{c}^{\dagger}_{i,m} \hat{c}_{j,m}-\sum_{i,m,m'}(\bm{F}(t)\cdot \bm{r}_{i}^{m'm})\hat{c}^{\dagger}_{i,m'} \hat{c}_{i,m}+{1\over 2}\sum_{i,m}U_{i}^{m}\hat{n}_{i,m}(\hat{n}_{i,m}-1)\nonumber\\
&=&-\sum_{ i,j ,m} \hbar t_{ij}^{m} \hat{c}^{\dagger}_{i,m} \hat{c}_{j,m}-{\hbar \Omega z_0}\sum_{i,m,m'}[\cos(\Omega t)\frac{x_{i}^{m'm}}{a}-\sin(\Omega t)\frac{y_{i}^{m'm}}{a}]\hat{c}^{\dagger}_{i,m'} \hat{c}_{i,m}+{1\over 2}\sum_{i,m}U_{i}^{m}\hat{n}_{i,m}(\hat{n}_{i,m}-1).
\end{eqnarray} 
By introducing a unitary transformation
\begin{equation}
\hat{U}_{c}(t)=\exp[{i \over \hbar}\sum_{i,m}(-m_{a} \bm{\dot{r}}_l(t) \cdot \bm{r}_{i}^{m}) \hat{c}_{i,m}^{\dagger} \hat{c}_{i,m}],
\end{equation}
the Hamiltonian is converted to
\begin{eqnarray}
\hat{H}_{c}'(t)&=&-\hbar \sum_{i,j,m} t_{ij}^{m}\ \hat{c}^{\dagger}_{i,m} \hat{c}_{j,m} e^{{i\over\hbar}m_{a} A\Omega \rho_{ij}\sin(\Omega t+\phi_{ij})}+{1\over 2}\sum_{i,m}U_{i}^{m}\hat{n}_{i,m}(\hat{n}_{i,m}-1)\nonumber\\
&-&{\hbar \Omega z_0}\sum_{i,m \neq m'}(\cos(\Omega t)\frac{x_{i}^{m' m}}{a}-\sin(\Omega t)\frac{y_{i}^{m'm}}{a})\hat{c}^{\dagger}_{i,m'}\hat{c}_{i,m}e^{{i \over \hbar}[m_{a} \bm{\dot{r}}_l(t) \cdot (\bm{r}_{i}^{m'}-\bm{r}_{i}^{m})]} .
\label{eqn:HC}
\end{eqnarray}
\end{widetext}
Based on the symmetry of honeycomb lattices, the centers of Wannier functions $w_{m}(\bm{r}-\bm{R}_{i})$ along $y$-direction for all bands are the same as the $y$-center of the lattice site $i$. However, the centers of Wannier functions $w_{m}(\bm{r}-\bm{R}_{i})$ along $x$-direction are not the same as the $x$-center of the lattice site. 
 We define the dimensionless parameters 
$\eta_{m'm}^{x}={x_{i}^{m'm}\over a}$ and $\eta_{m'm}^{y}={y_{i}^{m'm}\over a}$ to further simplify $\hat{H}_c'(t)$ to
\begin{eqnarray}\label{eq:HC22}
&\hat{H}_{c}&'(t)=-\hbar \sum_{i,j ,m} t_{ij}^{m}\ \hat{c}^{\dagger}_{i,m} \hat{c}_{j,m} e^{{i}z_0 \sin(\Omega t+\phi_{ij})}-{\hbar \Omega z_0}\sum_{i,m \neq m'}\nonumber\\
&&[\cos(\Omega t) \eta_{m'm}^{x} -\sin(\Omega t)\eta_{m'm}^{y}]\hat{c}^{\dagger}_{i,m'}\hat{c}_{i,m}e^{i z_{0} {\Delta l_{i}^{m'm}}\sin(\Omega t)}\nonumber \\
&&\ \ +{1\over 2}\sum_{i,m}U_{i}^{m}\hat{n}_{i,m}(\hat{n}_{i,m}-1).
\end{eqnarray}
Here $\Delta l_{i}^{m'm}={x_{i}^{m'}-x_{i}^{m}\over a} $ is characterizing the  difference between the centers of $m$ and $m'$ Wannier functions.

According to the Floquet theory, the solution of Schr\"{o}dinger equation $i\hbar d_{t}|\psi(t)\rangle=\hat{H}_{c}'(t)|\psi(t)\rangle$ has a form of $|\psi_\nu(t)\rangle=|u_{\nu}(t)\rangle e^{-{i\over \hbar}t\epsilon_{\nu}}$, where $|u_{\nu}(t)\rangle$ is a periodic function with a period of $T=2\pi/\Omega$. $|\psi(t)_{\nu}\rangle$ is called the Floquet state, $|u_{\nu}(t)\rangle$ is the Floquet mode, and $\epsilon_{\nu}$ is the quasienergy. $\vert\psi_\nu(t)\rangle$ is also an eigenstate of the time-evolution operator in one period $T$, \textit{i.e.}, 
\begin{equation}
\hat{U}(t_{0}+T,t_{0})\ |\psi_{\nu}(t_0)\rangle=e^{-{i\over\hbar}T\epsilon_{\nu}} |\psi_{\nu}(t_0)\rangle,
\end{equation}
where $\hat{U}(t_{0}+T,t_{0})$ denotes the time-evolution operator from $t_{0}$ to $t_{0}+T$ and the eigenvalue $e^{-{i\over\hbar}T\epsilon_{\nu}}$ does not depend on the start time $t_0$. By solving the eigenvalue problem of the time-evolution
operator, the phase factor $e^{-{i\over\hbar}T\epsilon_{\nu}}$ and the Floquet state $|\psi_{\nu}(t)\rangle$ are uniquely defined, while the corresponding quasienergies and Floquet modes are not unique. A Floquet state can be written as $|\psi_{\nu}(t)\rangle=|u_{\nu n_{p}}(t)\rangle e^{-{i\over\hbar}t\epsilon_{\nu n_{p}}}$, where $\epsilon_{\nu n_{p}}=\epsilon_{\nu}+n_{p}\hbar \Omega$ and $|u_{\nu n_{p}}(t)\rangle=|u_{\nu}(t)\rangle e^{i n_{p}\Omega t}$. For a particular $\nu$, there are a series of orthogonal Floquet modes $|u_{\nu n_{p}}(t)\rangle$ with  quasienergies $\epsilon_{\nu n_{p}}$.
The quasienergies and the Floquet modes are the eigenstates and eigenenergies of quasienergy operator $\hat{Q}(t)=\hat{H}_{c}'(t)-i\hbar d_{t}$, \textit{i.e.},
\begin{equation}
\hat{Q}|u_{\nu n_{p}}\rangle\rangle=\epsilon_{\nu n_{p}}|u_{\nu n_{p}}\rangle\rangle.
\end{equation}
Here the time-dependent state $|u(t)\rangle$ with a period $T$ is written as a double-ket $|u\rangle\rangle$. The scalar product for such a state is given by $\langle\langle u|v\rangle\rangle={1\over T}\int_{0}^{T} dt\ \langle u(t)|v(t)\rangle $. Similar to spatially periodic Hamiltonians, one can fix all quasienergies in the same interval of width $\hbar \omega$, called a Brillouin zone. Therefore, all Floquet states $|\psi_{\nu}(t)\rangle$ can be constructed from the Floquet modes whose quasienergies lie in a single Brillouin zone.

For the driven optical lattices, a useful set of bases is
\begin{equation}
|m,i,n_{p}\rangle\rangle=\hat{c}_{i,m}^{\dagger}|vac\rangle e^{i n_{p}\Omega t}.
\end{equation}
Here $n_{p}$ is the number of Floquet photons. Then the matrix elements of the quasienergy operator $\hat{Q}$ is 
\begin{equation}
\langle \langle m',i',n'_{p}|\hat{Q}|m,i,n_{p}\rangle\rangle = \langle m',i'|\hat{H}'_{c,n'_{p}-n_{p}}+\delta_{n'_{p}n_{p}}n_{p}\hbar\Omega|m,i \rangle,
\end{equation}
 where $\hat{H}'_{c,n}$ is obtained by the Fourier transformation of $\hat{H}'_{c}(t)=\sum_n \hat{H}'_{c,s}e^{is\Omega t}$ with a form of
\begin{eqnarray}
&\hat{H}&'_{c,s}=-\sum_{i,j ,m}t_{ij}^{m}J_{s}(z_{0})e^{is\phi_{ij}} \hat{c}_{i,m}^{\dagger}\hat{c}_{j,m}-\hbar \Omega z_{0}\sum_{i,m\neq m'}\nonumber\\
&&\left[\eta_{m'm}^{+}J_{s-1}(z_{0}\Delta l_{i}^{m'm})+\eta_{m'm}^{-}J_{s+1}(z_{0}\Delta l_{i}^{m'm})\right]\hat{c}^{\dagger}_{i,m'}\hat{c}_{i,m}.\nonumber\\
&&\ \  +{1\over 2}\delta_{s0}\sum_{i,m}U_{i}^{m}\hat{n}_{i,m}(\hat{n}_{i,m}-1)
\end{eqnarray}
Here $\eta_{m'm}^{+}$ ($\eta_{m'm}^{-}$) corresponds to ${\eta^{x}_{m'm}+i\eta^{y}_{m'm}\over 2}$ (${\eta^{x}_{m'm}-i\eta^{y}_{m'm}\over 2}$).
In Table~\ref{tab:Floquet}, we list the related numeric values of $\eta$ and $\Delta l$ for $s$- and $p$- bands.

Then we will estimate the resonant coupling strength of interband transitions via absorbing $\Delta n_p$ Floquet photons.
In our case, the Floquet photon energy
$\Omega={t}/0.108=0.287E_{r}$ and the band gap $-t_{ii}^{p}\approx 8.271E_{r}$, so it requires around 28 photons to excite a particle from $s$-band to $p$-band. For a $\Delta n_{p}$-photon transition process where $\Delta n_{p}$ is large enough, the coupling strength is 
\begin{eqnarray}
\langle\langle &P_x&,i,n_{p}|\hat{Q}|S ,i,n_{p}+\Delta n_{p}\rangle\rangle \sim J_{\Delta n_{p}-1}(z_{0}\Delta l_{i}^{m'm})\nonumber\\
&=& \sum_{k=0}^{\infty}{(-1)^{k}\over k! (k+\Delta n_{p})!}\left(\frac{z_{0}\Delta l_{i}^{m'm}}{2} \right)^{2k+\Delta n_{p}-1}\nonumber\\
&\sim&\frac{1}{\Delta n_{p}!}\left(\frac{z_{0}\Delta l_{i}^{m'm}}{2} \right)^{\Delta n_{p}-1}\nonumber\\
&\sim&\left(e \frac{z_{0}\Delta l_{i}^{m'm}}{2\Delta n_{p}} \right)^{\Delta n_{p}-1}=\left(\frac{z_{0}}{z_{th}}\right)^{\Delta n_{p}-1}.
\label{eqn:coupling}
\end{eqnarray}
Here we apply Stirling formula $(n)!=\sqrt{2\pi n} ({n\over e})^n$ to the factorial. Because $\Delta l_{i}^{m'm}$ is less than 1 between all bands, this suggests  $z_{th} > 2\frac{\Delta n_{p}}{e}$. When $z_0$ is smaller than $z_{th}$, the coupling strength is exponentially suppressed.

Besides absorbing $\Delta n_p$ photons directly, the $\Delta n_{p}$-order transition to the target state via $\Delta n_{p}-1$ 
intermediate states may also heat the system. 
For the $\Delta n_{p}$-order transition, the particle absorbs a single photon $\Delta n_{p}$ times. For each time the coupling strength is 
\begin{eqnarray}
\langle\langle P_x,i,n_{p}|&\hat{Q}&|S ,i,n_{p}+1\rangle\rangle \nonumber\\
&\approx& \hbar \Omega z_{0} \left({\eta^{x}_{m'm}+i\eta^{y}_{m'm}\over 2}\right) J_{0}(z_{0}\Delta l_{i}^{m'm})\nonumber\\
&<& \hbar \Omega z_{0} \left({\eta^{x}_{m'm}+i\eta^{y}_{m'm}\over 2}\right).
\end{eqnarray}
The coupling for the $\Delta n_{p}$-order process is $\langle\langle P_x,i,n_{p}|\hat{Q}|S ,i,n_{p}+1\rangle\rangle^{\Delta n_{p}}$ divided by the product of all energetic detuning of intermediate states. According to the discussion on high order transition processes in Ref.~\cite{Interband2016ZNTA,Multiphoton2015PRA}, this product has the same order of magnitude as $\frac{1}{(\Delta n_{p}-1)!}(\hbar \Omega)^{\Delta n_{p}-1}$. Therefore, the coupling term for a $\Delta n_{p}$-order process also behaves as $\left( {z_{0} \Delta l_{i}^{m'm}\over z_{th}}\right)^{\Delta n_{p}-1}$.  For a harmonic trap, the dipole matrix element is nonzero only for two states whose difference of the vibrational energy level is 1. Based on Table~\ref{tab:Floquet}, the dimensionless dipole matrix elements are less than 1, 
so the threshold $z_{th}$ for $\Delta n_{p}$-order process is larger than $\Delta n_{p}-1 \over e$.   

\begin{table}
\caption{\label{tab:Floquet}
Numeric value relevant to Floquet heating. 
}
\begin{ruledtabular}
\begin{tabular}{cccc}
$\eta_{p_{x}s}^{x}$&0.1292&$\Delta l_{i}^{p_{x}s}$ &0.0891\\
$\eta_{p_{y}s}^{x}$&0&$\Delta l_{i}^{p_{y}s}$ &0 \\
$\eta_{p_{x}s}^{y}$&0&\\
$\eta_{p_{y}s}^{y}$&0.1292& \\
\end{tabular}
\end{ruledtabular}
\end{table}

In our case, the target value of the modulation parameter $z_0$ is 2.3, which is less than the threshold, so the interband excitation caused by Floquet modulations is exponentially suppressed and negligible.

In the derivation of Eq.~\ref{eqn:HamiltonianEff}, the on-site interaction is neglected. Although the modulation couple the single-occupation state $g^{\dagger}_{A}g^{\dagger}_{B}|vac\rangle$ to the doublon state $g_{A}^{\dagger}g_{A}^{\dagger}|vac\rangle$, the coupling strength is suppressed by $J_{\Delta n_{p}}(z_{0})$ where $\Delta n_{p}$ is over 100. Therefore, it's safe to apply single-occupation state space under the Floquet modulation. 
 
\section{Many-body Hamiltonian}\label{App:Hamiltonian}

\begin{figure}
\centering
\includegraphics[width=0.5\textwidth]{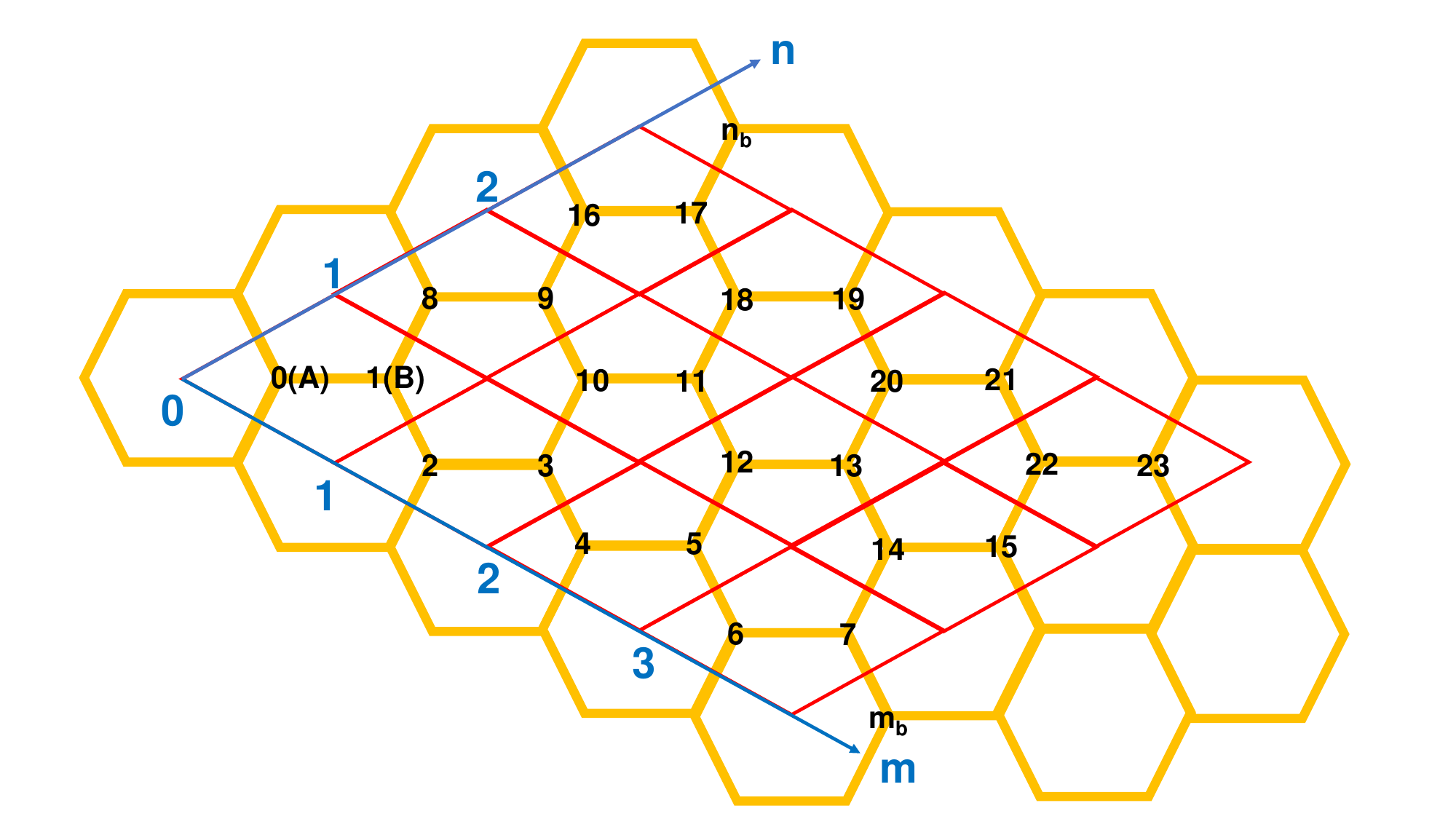}
\caption{24 lattice sites for exact diagonalization and the enumeration rule.}
\label{fig:LatticeNum}
\end{figure}

In this section, we use the case of 24 lattice sites ($3\times4\times2$, see Fig.~\ref{fig:LatticeNum}) and 6 particles as an example to illustrate how to enumerate Fock states and write down the many-body Hamiltonian for the exact- diagonalization calculation.

First, we give all sites a serial number, which are illustrated in Fig.~\ref{fig:LatticeNum}. $c_{n}^{\dagger}$ ($c_{n}$) is the creation (annihilation) operator of a boson on the $n$-th site, where $n=0,1,...,23$. For hard-core bosons, the basis is formed by $\{c_{i_{6}}^{\dagger}c^{\dagger}_{i_{5}}c^{\dagger}_{i_{4}}c^{\dagger}_{i_{3}}c^{\dagger}_{i_{2}}c^{\dagger}_{i_{1}}|\textrm{vac}\rangle\}$ where ${i_{1}< i_{2} < i_{3}< i_{4} < i_{5} < i_{6}}$ and $|\textrm{vac}\rangle$ is the vacuum state. The total number of base vectors is a binomial coefficient of sites and particles, which is 134596 in our case. We give all the occupation structure a serial number from 1 to 134596. The mapping rule is as follows
\begin{eqnarray}
\{c^{\dagger}_{i_{6}}c^{\dagger}_{i_{5}}c^{\dagger}_{i_{4}}c^{\dagger}_{i_{3}}c^{\dagger}_{i_{2}}c^{\dagger}_{i_{1}}|\textrm{vac}\rangle\}_{i_{1}< i_{2} < i_{3}< i_{4} < i_{5} < i_{6}}:& & \longrightarrow \nonumber \\
 \sum_{n_1=24-i_1}^{23}\sum_{n_2=24-i_2}^{22-i_1}\sum_{n_3=24-i_3}^{22-i_2}\sum_{n_4=24-i_4}^{22-i_3}\sum_{n_5=24-i_5}^{22-i_4}& &\nonumber\\
{n_1 \choose 5}{n_2 \choose 4}{n_3 \choose 3}{n_4 \choose 2}{n_5 \choose 1}+n_6-n_5
\end{eqnarray}

\begin{figure}
\centering
\includegraphics[width=0.5\textwidth]{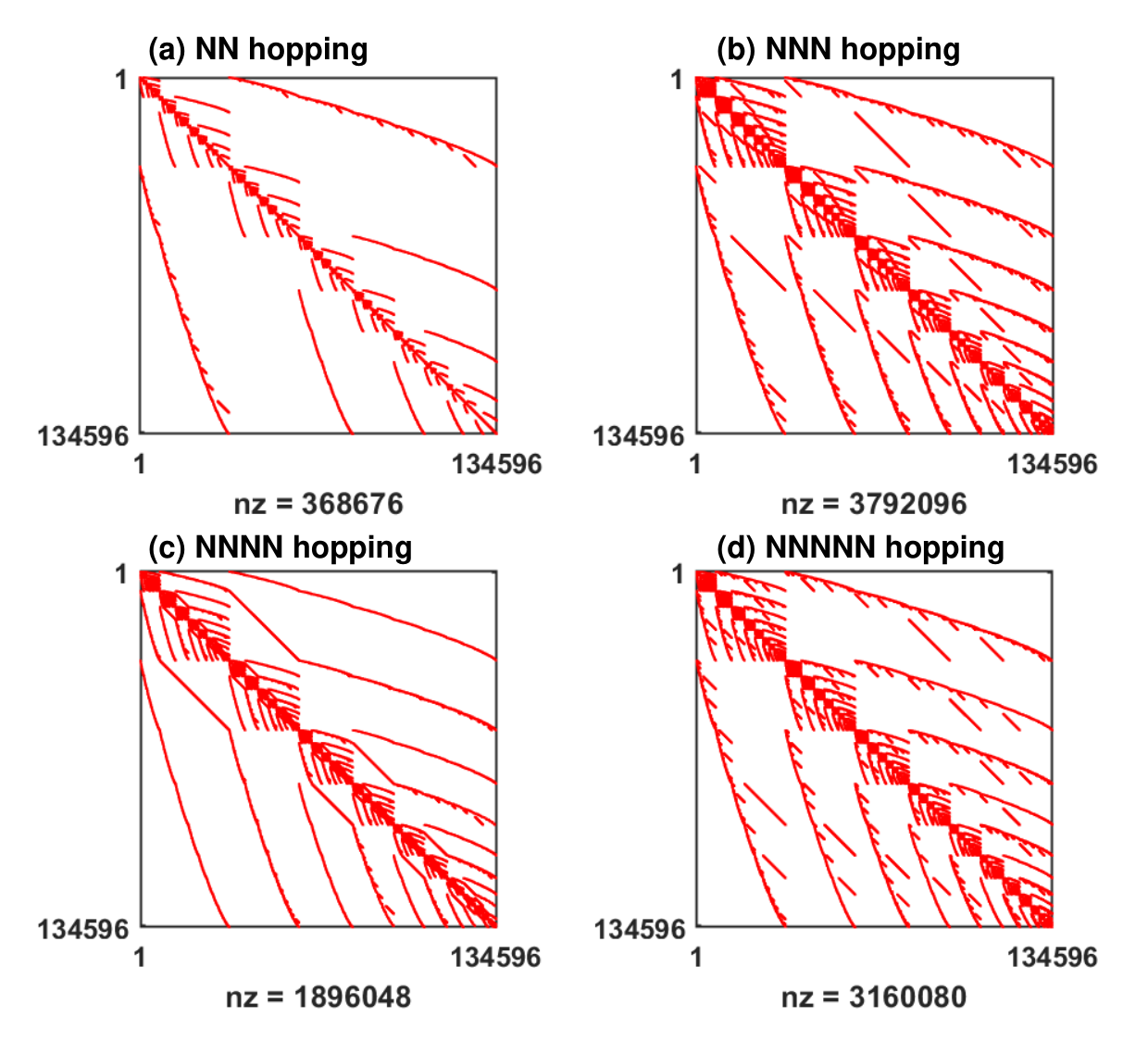}
\caption{(a-d) The non-zero elements of the NN, NNN, NNNN, NNNNN contributions are marked respectively. $nz$ is the number of non-zero elements.}
\label{fig:Hamiltonian}
\end{figure}

Besides, the twisted boundary condition should also be satisfied. To achieve this boundary conditions, the creation operators should satisfy $c^{\dagger}_{m_b}|\textrm{vac}\rangle=e^{i\theta_{x}}c^{\dagger}_0|\textrm{vac}\rangle$ and $c^{\dagger}_{n_b}|\textrm{vac}\rangle=e^{i\theta_{y}}c^{\dagger}_0|\textrm{vac}\rangle$ where $n_b$ and $m_b$ are positions outside the zone in Fig.~\ref{fig:LatticeNum}. 
Based on the twisted boundary condition, we map the tunneling out of this region back into the region of interest. We write out the many-body Hamiltonian based on this enumeration rule. To visualize the Hamiltonian, we mark the states connected by NN, NNN, NNNN, NNNN hopping terms respectively in Fig.~\ref{fig:Hamiltonian}.

\end{appendix}


\begin{thebibliography}{49}%
\makeatletter
\providecommand \@ifxundefined [1]{%
 \@ifx{#1\undefined}
}%
\providecommand \@ifnum [1]{%
 \ifnum #1\expandafter \@firstoftwo
 \else \expandafter \@secondoftwo
 \fi
}%
\providecommand \@ifx [1]{%
 \ifx #1\expandafter \@firstoftwo
 \else \expandafter \@secondoftwo
 \fi
}%
\providecommand \natexlab [1]{#1}%
\providecommand \enquote  [1]{``#1''}%
\providecommand \bibnamefont  [1]{#1}%
\providecommand \bibfnamefont [1]{#1}%
\providecommand \citenamefont [1]{#1}%
\providecommand \href@noop [0]{\@secondoftwo}%
\providecommand \href [0]{\begingroup \@sanitize@url \@href}%
\providecommand \@href[1]{\@@startlink{#1}\@@href}%
\providecommand \@@href[1]{\endgroup#1\@@endlink}%
\providecommand \@sanitize@url [0]{\catcode `\\12\catcode `\$12\catcode
  `\&12\catcode `\#12\catcode `\^12\catcode `\_12\catcode `\%12\relax}%
\providecommand \@@startlink[1]{}%
\providecommand \@@endlink[0]{}%
\providecommand \url  [0]{\begingroup\@sanitize@url \@url }%
\providecommand \@url [1]{\endgroup\@href {#1}{\urlprefix }}%
\providecommand \urlprefix  [0]{URL }%
\providecommand \Eprint [0]{\href }%
\providecommand \doibase [0]{http://dx.doi.org/}%
\providecommand \selectlanguage [0]{\@gobble}%
\providecommand \bibinfo  [0]{\@secondoftwo}%
\providecommand \bibfield  [0]{\@secondoftwo}%
\providecommand \translation [1]{[#1]}%
\providecommand \BibitemOpen [0]{}%
\providecommand \bibitemStop [0]{}%
\providecommand \bibitemNoStop [0]{.\EOS\space}%
\providecommand \EOS [0]{\spacefactor3000\relax}%
\providecommand \BibitemShut  [1]{\csname bibitem#1\endcsname}%
\let\auto@bib@innerbib\@empty
\bibitem [{\citenamefont {Stormer}\ \emph {et~al.}(1999)\citenamefont
  {Stormer}, \citenamefont {Tsui},\ and\ \citenamefont
  {Gossard}}]{Stormer1999RMP}%
  \BibitemOpen
  \bibfield  {author} {\bibinfo {author} {\bibfnamefont {Horst~L.}\
  \bibnamefont {Stormer}}, \bibinfo {author} {\bibfnamefont {Daniel~C.}\
  \bibnamefont {Tsui}}, \ and\ \bibinfo {author} {\bibfnamefont {Arthur~C.}\
  \bibnamefont {Gossard}},\ }\bibfield  {title} {\enquote {\bibinfo {title}
  {The fractional quantum Hall effect},}\ }\href {\doibase
  10.1103/RevModPhys.71.S298} {\bibfield  {journal} {\bibinfo  {journal} {Rev.
  Mod. Phys.}\ }\textbf {\bibinfo {volume} {71}},\ \bibinfo {pages}
  {S298--S305} (\bibinfo {year} {1999})}\BibitemShut {NoStop}%
\bibitem [{\citenamefont {Cooper}(2020)}]{CooperBook}%
  \BibitemOpen
  \bibfield  {author} {\bibinfo {author} {\bibfnamefont {N.R.}\ \bibnamefont
  {Cooper}},\ }\enquote {\bibinfo {title} {Fractional quantum Hall states of
  bosons: Properties and prospects for experimental realization},}\ in\ \href
  {\doibase 10.1142/9789811217494_0010} {\emph {\bibinfo {booktitle}
  {Fractional Quantum Hall Effects: New Developments}}}\ (\bibinfo  {publisher}
  {World Scientific Press},\ \bibinfo {year} {2020})\ Chap.~\bibinfo {chapter}
  {10}, pp.\ \bibinfo {pages} {487--521}\BibitemShut {NoStop}%
\bibitem [{\citenamefont {Wen}(2017)}]{Wen2017RMP}%
  \BibitemOpen
  \bibfield  {author} {\bibinfo {author} {\bibfnamefont {Xiao-Gang}\
  \bibnamefont {Wen}},\ }\bibfield  {title} {\enquote {\bibinfo {title}
  {Colloquium: Zoo of quantum-topological phases of matter},}\ }\href {\doibase
  10.1103/RevModPhys.89.041004} {\bibfield  {journal} {\bibinfo  {journal}
  {Rev. Mod. Phys.}\ }\textbf {\bibinfo {volume} {89}},\ \bibinfo {pages}
  {041004} (\bibinfo {year} {2017})}\BibitemShut {NoStop}%
\bibitem [{\citenamefont {Bloch}\ \emph {et~al.}(2008)\citenamefont {Bloch},
  \citenamefont {Dalibard},\ and\ \citenamefont {Zwerger}}]{Bloch2008RMP}%
  \BibitemOpen
  \bibfield  {author} {\bibinfo {author} {\bibfnamefont {Immanuel}\
  \bibnamefont {Bloch}}, \bibinfo {author} {\bibfnamefont {Jean}\ \bibnamefont
  {Dalibard}}, \ and\ \bibinfo {author} {\bibfnamefont {Wilhelm}\ \bibnamefont
  {Zwerger}},\ }\bibfield  {title} {\enquote {\bibinfo {title} {Many-body
  physics with ultracold gases},}\ }\href {\doibase 10.1103/RevModPhys.80.885}
  {\bibfield  {journal} {\bibinfo  {journal} {Rev. Mod. Phys.}\ }\textbf
  {\bibinfo {volume} {80}},\ \bibinfo {pages} {885--964} (\bibinfo {year}
  {2008})}\BibitemShut {NoStop}%
\bibitem [{\citenamefont {Cooper}\ \emph {et~al.}(2019)\citenamefont {Cooper},
  \citenamefont {Dalibard},\ and\ \citenamefont {Spielman}}]{Cooper2019RMP}%
  \BibitemOpen
  \bibfield  {author} {\bibinfo {author} {\bibfnamefont {N.~R.}\ \bibnamefont
  {Cooper}}, \bibinfo {author} {\bibfnamefont {J.}~\bibnamefont {Dalibard}}, \
  and\ \bibinfo {author} {\bibfnamefont {I.~B.}\ \bibnamefont {Spielman}},\
  }\bibfield  {title} {\enquote {\bibinfo {title} {Topological bands for
  ultracold atoms},}\ }\href {\doibase 10.1103/RevModPhys.91.015005} {\bibfield
   {journal} {\bibinfo  {journal} {Rev. Mod. Phys.}\ }\textbf {\bibinfo
  {volume} {91}},\ \bibinfo {pages} {015005} (\bibinfo {year}
  {2019})}\BibitemShut {NoStop}%
\bibitem [{\citenamefont {S\o{}rensen}\ \emph {et~al.}(2005)\citenamefont
  {S\o{}rensen}, \citenamefont {Demler},\ and\ \citenamefont
  {Lukin}}]{Lukin2005PRL}%
  \BibitemOpen
  \bibfield  {author} {\bibinfo {author} {\bibfnamefont {Anders~S.}\
  \bibnamefont {S\o{}rensen}}, \bibinfo {author} {\bibfnamefont {Eugene}\
  \bibnamefont {Demler}}, \ and\ \bibinfo {author} {\bibfnamefont {Mikhail~D.}\
  \bibnamefont {Lukin}},\ }\bibfield  {title} {\enquote {\bibinfo {title}
  {Fractional quantum Hall states of atoms in optical lattices},}\ }\href
  {\doibase 10.1103/PhysRevLett.94.086803} {\bibfield  {journal} {\bibinfo
  {journal} {Phys. Rev. Lett.}\ }\textbf {\bibinfo {volume} {94}},\ \bibinfo
  {pages} {086803} (\bibinfo {year} {2005})}\BibitemShut {NoStop}%
\bibitem [{\citenamefont {Hafezi}\ \emph {et~al.}(2007)\citenamefont {Hafezi},
  \citenamefont {S\o{}rensen}, \citenamefont {Demler},\ and\ \citenamefont
  {Lukin}}]{hafezi2007fractional}%
  \BibitemOpen
  \bibfield  {author} {\bibinfo {author} {\bibfnamefont {M.}~\bibnamefont
  {Hafezi}}, \bibinfo {author} {\bibfnamefont {A.~S.}\ \bibnamefont
  {S\o{}rensen}}, \bibinfo {author} {\bibfnamefont {E.}~\bibnamefont {Demler}},
  \ and\ \bibinfo {author} {\bibfnamefont {M.~D.}\ \bibnamefont {Lukin}},\
  }\bibfield  {title} {\enquote {\bibinfo {title} {Fractional quantum Hall
  effect in optical lattices},}\ }\href {\doibase 10.1103/PhysRevA.76.023613}
  {\bibfield  {journal} {\bibinfo  {journal} {Phys. Rev. A}\ }\textbf {\bibinfo
  {volume} {76}},\ \bibinfo {pages} {023613} (\bibinfo {year}
  {2007})}\BibitemShut {NoStop}%
\bibitem [{\citenamefont {Cooper}\ and\ \citenamefont
  {Dalibard}(2013{\natexlab{a}})}]{Cooper2013PRL}%
  \BibitemOpen
  \bibfield  {author} {\bibinfo {author} {\bibfnamefont {Nigel~R.}\
  \bibnamefont {Cooper}}\ and\ \bibinfo {author} {\bibfnamefont {Jean}\
  \bibnamefont {Dalibard}},\ }\bibfield  {title} {\enquote {\bibinfo {title}
  {Reaching fractional quantum Hall states with optical flux lattices},}\
  }\href {\doibase 10.1103/PhysRevLett.110.185301} {\bibfield  {journal}
  {\bibinfo  {journal} {Phys. Rev. Lett.}\ }\textbf {\bibinfo {volume} {110}},\
  \bibinfo {pages} {185301} (\bibinfo {year} {2013}{\natexlab{a}})}\BibitemShut
  {NoStop}%
\bibitem [{\citenamefont {Wang}\ \emph {et~al.}(2011)\citenamefont {Wang},
  \citenamefont {Gu}, \citenamefont {Gong},\ and\ \citenamefont
  {Sheng}}]{2011Fractional}%
  \BibitemOpen
  \bibfield  {author} {\bibinfo {author} {\bibfnamefont {Yi-Fei}\ \bibnamefont
  {Wang}}, \bibinfo {author} {\bibfnamefont {Zheng-Cheng}\ \bibnamefont {Gu}},
  \bibinfo {author} {\bibfnamefont {Chang-De}\ \bibnamefont {Gong}}, \ and\
  \bibinfo {author} {\bibfnamefont {D.~N.}\ \bibnamefont {Sheng}},\ }\bibfield
  {title} {\enquote {\bibinfo {title} {Fractional quantum Hall effect of
  hard-core bosons in topological flat bands},}\ }\href {\doibase
  10.1103/PhysRevLett.107.146803} {\bibfield  {journal} {\bibinfo  {journal}
  {Phys. Rev. Lett.}\ }\textbf {\bibinfo {volume} {107}},\ \bibinfo {pages}
  {146803} (\bibinfo {year} {2011})}\BibitemShut {NoStop}%
\bibitem [{\citenamefont {Neupert}\ \emph {et~al.}(2011)\citenamefont
  {Neupert}, \citenamefont {Santos}, \citenamefont {Chamon},\ and\
  \citenamefont {Mudry}}]{neupert2011fractional}%
  \BibitemOpen
  \bibfield  {author} {\bibinfo {author} {\bibfnamefont {Titus}\ \bibnamefont
  {Neupert}}, \bibinfo {author} {\bibfnamefont {Luiz}\ \bibnamefont {Santos}},
  \bibinfo {author} {\bibfnamefont {Claudio}\ \bibnamefont {Chamon}}, \ and\
  \bibinfo {author} {\bibfnamefont {Christopher}\ \bibnamefont {Mudry}},\
  }\bibfield  {title} {\enquote {\bibinfo {title} {Fractional quantum Hall
  states at zero magnetic field},}\ }\href {\doibase
  10.1103/PhysRevLett.106.236804} {\bibfield  {journal} {\bibinfo  {journal}
  {Phys. Rev. Lett.}\ }\textbf {\bibinfo {volume} {106}},\ \bibinfo {pages}
  {236804} (\bibinfo {year} {2011})}\BibitemShut {NoStop}%
\bibitem [{\citenamefont {Hudomal}\ \emph {et~al.}(2019)\citenamefont
  {Hudomal}, \citenamefont {Regnault},\ and\ \citenamefont
  {Vasi\ifmmode~\acute{c}\else \'{c}\fi{}}}]{PhysRevA.100.053624}%
  \BibitemOpen
  \bibfield  {author} {\bibinfo {author} {\bibfnamefont {Ana}\ \bibnamefont
  {Hudomal}}, \bibinfo {author} {\bibfnamefont {Nicolas}\ \bibnamefont
  {Regnault}}, \ and\ \bibinfo {author} {\bibfnamefont {Ivana}\ \bibnamefont
  {Vasi\ifmmode~\acute{c}\else \'{c}\fi{}}},\ }\bibfield  {title} {\enquote
  {\bibinfo {title} {Bosonic fractional quantum Hall states in driven optical
  lattices},}\ }\href {\doibase 10.1103/PhysRevA.100.053624} {\bibfield
  {journal} {\bibinfo  {journal} {Phys. Rev. A}\ }\textbf {\bibinfo {volume}
  {100}},\ \bibinfo {pages} {053624} (\bibinfo {year} {2019})}\BibitemShut
  {NoStop}%
\bibitem [{\citenamefont {Sun}\ \emph {et~al.}(2011)\citenamefont {Sun},
  \citenamefont {Gu}, \citenamefont {Katsura},\ and\ \citenamefont
  {Das~Sarma}}]{Sun2011PRL}%
  \BibitemOpen
  \bibfield  {author} {\bibinfo {author} {\bibfnamefont {Kai}\ \bibnamefont
  {Sun}}, \bibinfo {author} {\bibfnamefont {Zhengcheng}\ \bibnamefont {Gu}},
  \bibinfo {author} {\bibfnamefont {Hosho}\ \bibnamefont {Katsura}}, \ and\
  \bibinfo {author} {\bibfnamefont {S.}~\bibnamefont {Das~Sarma}},\ }\bibfield
  {title} {\enquote {\bibinfo {title} {Nearly flatbands with nontrivial
  topology},}\ }\href {\doibase 10.1103/PhysRevLett.106.236803} {\bibfield
  {journal} {\bibinfo  {journal} {Phys. Rev. Lett.}\ }\textbf {\bibinfo
  {volume} {106}},\ \bibinfo {pages} {236803} (\bibinfo {year}
  {2011})}\BibitemShut {NoStop}%
\bibitem [{\citenamefont {Tang}\ \emph {et~al.}(2011)\citenamefont {Tang},
  \citenamefont {Mei},\ and\ \citenamefont {Wen}}]{Tang2011PRL}%
  \BibitemOpen
  \bibfield  {author} {\bibinfo {author} {\bibfnamefont {Evelyn}\ \bibnamefont
  {Tang}}, \bibinfo {author} {\bibfnamefont {Jia-Wei}\ \bibnamefont {Mei}}, \
  and\ \bibinfo {author} {\bibfnamefont {Xiao-Gang}\ \bibnamefont {Wen}},\
  }\bibfield  {title} {\enquote {\bibinfo {title} {High-temperature fractional
  quantum Hall states},}\ }\href {\doibase 10.1103/PhysRevLett.106.236802}
  {\bibfield  {journal} {\bibinfo  {journal} {Phys. Rev. Lett.}\ }\textbf
  {\bibinfo {volume} {106}},\ \bibinfo {pages} {236802} (\bibinfo {year}
  {2011})}\BibitemShut {NoStop}%
\bibitem [{\citenamefont {Grushin}\ \emph {et~al.}(2014)\citenamefont
  {Grushin}, \citenamefont {G\'omez-Le\'on},\ and\ \citenamefont
  {Neupert}}]{Grushin2014PRL}%
  \BibitemOpen
  \bibfield  {author} {\bibinfo {author} {\bibfnamefont {Adolfo~G.}\
  \bibnamefont {Grushin}}, \bibinfo {author} {\bibfnamefont {\'Alvaro}\
  \bibnamefont {G\'omez-Le\'on}}, \ and\ \bibinfo {author} {\bibfnamefont
  {Titus}\ \bibnamefont {Neupert}},\ }\bibfield  {title} {\enquote {\bibinfo
  {title} {Floquet fractional Chern insulators},}\ }\href {\doibase
  10.1103/PhysRevLett.112.156801} {\bibfield  {journal} {\bibinfo  {journal}
  {Phys. Rev. Lett.}\ }\textbf {\bibinfo {volume} {112}},\ \bibinfo {pages}
  {156801} (\bibinfo {year} {2014})}\BibitemShut {NoStop}%
\bibitem [{\citenamefont {Cooper}\ and\ \citenamefont
  {Dalibard}(2013{\natexlab{b}})}]{PhysRevLett.110.185301}%
  \BibitemOpen
  \bibfield  {author} {\bibinfo {author} {\bibfnamefont {Nigel~R.}\
  \bibnamefont {Cooper}}\ and\ \bibinfo {author} {\bibfnamefont {Jean}\
  \bibnamefont {Dalibard}},\ }\bibfield  {title} {\enquote {\bibinfo {title}
  {Reaching fractional quantum Hall states with optical flux lattices},}\
  }\href {\doibase 10.1103/PhysRevLett.110.185301} {\bibfield  {journal}
  {\bibinfo  {journal} {Phys. Rev. Lett.}\ }\textbf {\bibinfo {volume} {110}},\
  \bibinfo {pages} {185301} (\bibinfo {year} {2013}{\natexlab{b}})}\BibitemShut
  {NoStop}%
\bibitem [{\citenamefont {Miyake}\ \emph {et~al.}(2013)\citenamefont {Miyake},
  \citenamefont {Siviloglou}, \citenamefont {Kennedy}, \citenamefont {Burton},\
  and\ \citenamefont {Ketterle}}]{miyake2013realizing}%
  \BibitemOpen
  \bibfield  {author} {\bibinfo {author} {\bibfnamefont {Hirokazu}\
  \bibnamefont {Miyake}}, \bibinfo {author} {\bibfnamefont {Georgios~A.}\
  \bibnamefont {Siviloglou}}, \bibinfo {author} {\bibfnamefont {Colin~J.}\
  \bibnamefont {Kennedy}}, \bibinfo {author} {\bibfnamefont {William~Cody}\
  \bibnamefont {Burton}}, \ and\ \bibinfo {author} {\bibfnamefont {Wolfgang}\
  \bibnamefont {Ketterle}},\ }\bibfield  {title} {\enquote {\bibinfo {title}
  {Realizing the Harper Hamiltonian with laser-assisted tunneling in optical
  lattices},}\ }\href {\doibase 10.1103/PhysRevLett.111.185302} {\bibfield
  {journal} {\bibinfo  {journal} {Phys. Rev. Lett.}\ }\textbf {\bibinfo
  {volume} {111}},\ \bibinfo {pages} {185302} (\bibinfo {year}
  {2013})}\BibitemShut {NoStop}%
\bibitem [{\citenamefont {Aidelsburger}\ \emph {et~al.}(2013)\citenamefont
  {Aidelsburger}, \citenamefont {Atala}, \citenamefont {Lohse}, \citenamefont
  {Barreiro}, \citenamefont {Paredes},\ and\ \citenamefont
  {Bloch}}]{Aidelsburger2013}%
  \BibitemOpen
  \bibfield  {author} {\bibinfo {author} {\bibfnamefont {M.}~\bibnamefont
  {Aidelsburger}}, \bibinfo {author} {\bibfnamefont {M.}~\bibnamefont {Atala}},
  \bibinfo {author} {\bibfnamefont {M.}~\bibnamefont {Lohse}}, \bibinfo
  {author} {\bibfnamefont {J.~T.}\ \bibnamefont {Barreiro}}, \bibinfo {author}
  {\bibfnamefont {B.}~\bibnamefont {Paredes}}, \ and\ \bibinfo {author}
  {\bibfnamefont {I.}~\bibnamefont {Bloch}},\ }\bibfield  {title} {\enquote
  {\bibinfo {title} {Realization of the Hofstadter Hamiltonian with ultracold
  atoms in optical lattices},}\ }\href {\doibase
  10.1103/PhysRevLett.111.185301} {\bibfield  {journal} {\bibinfo  {journal}
  {Phys. Rev. Lett.}\ }\textbf {\bibinfo {volume} {111}},\ \bibinfo {pages}
  {185301} (\bibinfo {year} {2013})}\BibitemShut {NoStop}%
\bibitem [{\citenamefont {Jotzu}\ \emph {et~al.}(2014)\citenamefont {Jotzu},
  \citenamefont {Messer}, \citenamefont {Desbuquois}, \citenamefont {Lebrat},
  \citenamefont {Uehlinger}, \citenamefont {Greif},\ and\ \citenamefont
  {Esslinger}}]{Jotzu2014}%
  \BibitemOpen
  \bibfield  {author} {\bibinfo {author} {\bibfnamefont {Gregor}\ \bibnamefont
  {Jotzu}}, \bibinfo {author} {\bibfnamefont {Michael}\ \bibnamefont {Messer}},
  \bibinfo {author} {\bibfnamefont {R{\'{e}}mi}\ \bibnamefont {Desbuquois}},
  \bibinfo {author} {\bibfnamefont {Martin}\ \bibnamefont {Lebrat}}, \bibinfo
  {author} {\bibfnamefont {Thomas}\ \bibnamefont {Uehlinger}}, \bibinfo
  {author} {\bibfnamefont {Daniel}\ \bibnamefont {Greif}}, \ and\ \bibinfo
  {author} {\bibfnamefont {Tilman}\ \bibnamefont {Esslinger}},\ }\bibfield
  {title} {\enquote {\bibinfo {title} {{Experimental realization of the
  topological Haldane model with ultracold fermions}},}\ }\href {\doibase
  10.1038/nature13915} {\bibfield  {journal} {\bibinfo  {journal} {Nature}\
  }\textbf {\bibinfo {volume} {515}},\ \bibinfo {pages} {237--240} (\bibinfo
  {year} {2014})}\BibitemShut {NoStop}%
\bibitem [{\citenamefont {Aidelsburger}\ \emph {et~al.}(2015)\citenamefont
  {Aidelsburger}, \citenamefont {Lohse}, \citenamefont {Schweizer},
  \citenamefont {Atala}, \citenamefont {Barreiro}, \citenamefont
  {Nascimb{\`e}ne}, \citenamefont {Cooper}, \citenamefont {Bloch},\ and\
  \citenamefont {Goldman}}]{aidelsburger2015measuring}%
  \BibitemOpen
  \bibfield  {author} {\bibinfo {author} {\bibfnamefont {Monika}\ \bibnamefont
  {Aidelsburger}}, \bibinfo {author} {\bibfnamefont {Michael}\ \bibnamefont
  {Lohse}}, \bibinfo {author} {\bibfnamefont {Christian}\ \bibnamefont
  {Schweizer}}, \bibinfo {author} {\bibfnamefont {Marcos}\ \bibnamefont
  {Atala}}, \bibinfo {author} {\bibfnamefont {Julio~T}\ \bibnamefont
  {Barreiro}}, \bibinfo {author} {\bibfnamefont {Sylvain}\ \bibnamefont
  {Nascimb{\`e}ne}}, \bibinfo {author} {\bibfnamefont {NR}~\bibnamefont
  {Cooper}}, \bibinfo {author} {\bibfnamefont {Immanuel}\ \bibnamefont
  {Bloch}}, \ and\ \bibinfo {author} {\bibfnamefont {Nathan}\ \bibnamefont
  {Goldman}},\ }\bibfield  {title} {\enquote {\bibinfo {title} {Measuring the
  Chern number of Hofstadter bands with ultracold bosonic atoms},}\ }\href@noop
  {} {\bibfield  {journal} {\bibinfo  {journal} {Nature Physics}\ }\textbf
  {\bibinfo {volume} {11}},\ \bibinfo {pages} {162--166} (\bibinfo {year}
  {2015})}\BibitemShut {NoStop}%
\bibitem [{\citenamefont {Aidelsburger}\ \emph {et~al.}(2011)\citenamefont
  {Aidelsburger}, \citenamefont {Atala}, \citenamefont {Nascimb\`ene},
  \citenamefont {Trotzky}, \citenamefont {Chen},\ and\ \citenamefont
  {Bloch}}]{Monika2011PRL}%
  \BibitemOpen
  \bibfield  {author} {\bibinfo {author} {\bibfnamefont {M.}~\bibnamefont
  {Aidelsburger}}, \bibinfo {author} {\bibfnamefont {M.}~\bibnamefont {Atala}},
  \bibinfo {author} {\bibfnamefont {S.}~\bibnamefont {Nascimb\`ene}}, \bibinfo
  {author} {\bibfnamefont {S.}~\bibnamefont {Trotzky}}, \bibinfo {author}
  {\bibfnamefont {Y.-A.}\ \bibnamefont {Chen}}, \ and\ \bibinfo {author}
  {\bibfnamefont {I.}~\bibnamefont {Bloch}},\ }\bibfield  {title} {\enquote
  {\bibinfo {title} {Experimental realization of strong effective magnetic
  fields in an optical lattice},}\ }\href {\doibase
  10.1103/PhysRevLett.107.255301} {\bibfield  {journal} {\bibinfo  {journal}
  {Phys. Rev. Lett.}\ }\textbf {\bibinfo {volume} {107}},\ \bibinfo {pages}
  {255301} (\bibinfo {year} {2011})}\BibitemShut {NoStop}%
\bibitem [{\citenamefont {Wintersperger}\ \emph {et~al.}(2020)\citenamefont
  {Wintersperger}, \citenamefont {Braun}, \citenamefont {{\"U}nal},
  \citenamefont {Eckardt}, \citenamefont {Di~Liberto}, \citenamefont {Goldman},
  \citenamefont {Bloch},\ and\ \citenamefont
  {Aidelsburger}}]{Monika2020realization}%
  \BibitemOpen
  \bibfield  {author} {\bibinfo {author} {\bibfnamefont {Karen}\ \bibnamefont
  {Wintersperger}}, \bibinfo {author} {\bibfnamefont {Christoph}\ \bibnamefont
  {Braun}}, \bibinfo {author} {\bibfnamefont {F~Nur}\ \bibnamefont {{\"U}nal}},
  \bibinfo {author} {\bibfnamefont {Andr{\'e}}\ \bibnamefont {Eckardt}},
  \bibinfo {author} {\bibfnamefont {Marco}\ \bibnamefont {Di~Liberto}},
  \bibinfo {author} {\bibfnamefont {Nathan}\ \bibnamefont {Goldman}}, \bibinfo
  {author} {\bibfnamefont {Immanuel}\ \bibnamefont {Bloch}}, \ and\ \bibinfo
  {author} {\bibfnamefont {Monika}\ \bibnamefont {Aidelsburger}},\ }\bibfield
  {title} {\enquote {\bibinfo {title} {Realization of an anomalous Floquet
  topological system with ultracold atoms},}\ }\href@noop {} {\bibfield
  {journal} {\bibinfo  {journal} {Nature Physics}\ }\textbf {\bibinfo {volume}
  {16}},\ \bibinfo {pages} {1058--1063} (\bibinfo {year} {2020})}\BibitemShut
  {NoStop}%
\bibitem [{\citenamefont {Beeler}\ \emph {et~al.}(2013)\citenamefont {Beeler},
  \citenamefont {Williams}, \citenamefont {Jimenez-Garcia}, \citenamefont
  {LeBlanc}, \citenamefont {Perry},\ and\ \citenamefont
  {Spielman}}]{beeler2013spin}%
  \BibitemOpen
  \bibfield  {author} {\bibinfo {author} {\bibfnamefont {M.~C.}\ \bibnamefont
  {Beeler}}, \bibinfo {author} {\bibfnamefont {R.~A.}\ \bibnamefont
  {Williams}}, \bibinfo {author} {\bibfnamefont {K.}~\bibnamefont
  {Jimenez-Garcia}}, \bibinfo {author} {\bibfnamefont {L.~J.}\ \bibnamefont
  {LeBlanc}}, \bibinfo {author} {\bibfnamefont {A.~R.}\ \bibnamefont {Perry}},
  \ and\ \bibinfo {author} {\bibfnamefont {I.~B.}\ \bibnamefont {Spielman}},\
  }\bibfield  {title} {\enquote {\bibinfo {title} {The spin Hall effect in a
  quantum gas},}\ }\href@noop {} {\bibfield  {journal} {\bibinfo  {journal}
  {Nature}\ }\textbf {\bibinfo {volume} {498}},\ \bibinfo {pages} {201--204}
  (\bibinfo {year} {2013})}\BibitemShut {NoStop}%
\bibitem [{\citenamefont {Stuhl}\ \emph {et~al.}(2015)\citenamefont {Stuhl},
  \citenamefont {Lu}, \citenamefont {Aycock}, \citenamefont {Genkina},\ and\
  \citenamefont {Spielman}}]{stuhl2015visualizing}%
  \BibitemOpen
  \bibfield  {author} {\bibinfo {author} {\bibfnamefont {B.K.}\ \bibnamefont
  {Stuhl}}, \bibinfo {author} {\bibfnamefont {H.-I.}\ \bibnamefont {Lu}},
  \bibinfo {author} {\bibfnamefont {L.M.}\ \bibnamefont {Aycock}}, \bibinfo
  {author} {\bibfnamefont {D.}~\bibnamefont {Genkina}}, \ and\ \bibinfo
  {author} {\bibfnamefont {I.B.}\ \bibnamefont {Spielman}},\ }\bibfield
  {title} {\enquote {\bibinfo {title} {Visualizing edge states with an atomic
  bose gas in the quantum Hall regime},}\ }\href@noop {} {\bibfield  {journal}
  {\bibinfo  {journal} {Science}\ }\textbf {\bibinfo {volume} {349}},\ \bibinfo
  {pages} {1514--1518} (\bibinfo {year} {2015})}\BibitemShut {NoStop}%
\bibitem [{\citenamefont {Lin}\ \emph {et~al.}(2009)\citenamefont {Lin},
  \citenamefont {Compton}, \citenamefont {Jim{\'e}nez-Garc{\'\i}a},
  \citenamefont {Porto},\ and\ \citenamefont {Spielman}}]{lin2009synthetic}%
  \BibitemOpen
  \bibfield  {author} {\bibinfo {author} {\bibfnamefont {Y.-J.}\ \bibnamefont
  {Lin}}, \bibinfo {author} {\bibfnamefont {Rob~L.}\ \bibnamefont {Compton}},
  \bibinfo {author} {\bibfnamefont {Karina}\ \bibnamefont
  {Jim{\'e}nez-Garc{\'\i}a}}, \bibinfo {author} {\bibfnamefont {James~V.}\
  \bibnamefont {Porto}}, \ and\ \bibinfo {author} {\bibfnamefont {Ian~B.}\
  \bibnamefont {Spielman}},\ }\bibfield  {title} {\enquote {\bibinfo {title}
  {Synthetic magnetic fields for ultracold neutral atoms},}\ }\href@noop {}
  {\bibfield  {journal} {\bibinfo  {journal} {Nature}\ }\textbf {\bibinfo
  {volume} {462}},\ \bibinfo {pages} {628--632} (\bibinfo {year}
  {2009})}\BibitemShut {NoStop}%
\bibitem [{\citenamefont {Haldane}(1988)}]{haldane1988model}%
  \BibitemOpen
  \bibfield  {author} {\bibinfo {author} {\bibfnamefont {F.~D.~M.}\
  \bibnamefont {Haldane}},\ }\bibfield  {title} {\enquote {\bibinfo {title}
  {Model for a quantum Hall effect without Landau levels: Condensed-matter
  realization of the "parity anomaly"},}\ }\href {\doibase
  10.1103/PhysRevLett.61.2015} {\bibfield  {journal} {\bibinfo  {journal}
  {Phys. Rev. Lett.}\ }\textbf {\bibinfo {volume} {61}},\ \bibinfo {pages}
  {2015--2018} (\bibinfo {year} {1988})}\BibitemShut {NoStop}%
\bibitem [{\citenamefont {Chin}\ \emph {et~al.}(2010)\citenamefont {Chin},
  \citenamefont {Grimm}, \citenamefont {Julienne},\ and\ \citenamefont
  {Tiesinga}}]{Chin2010RMP}%
  \BibitemOpen
  \bibfield  {author} {\bibinfo {author} {\bibfnamefont {Cheng}\ \bibnamefont
  {Chin}}, \bibinfo {author} {\bibfnamefont {Rudolf}\ \bibnamefont {Grimm}},
  \bibinfo {author} {\bibfnamefont {Paul}\ \bibnamefont {Julienne}}, \ and\
  \bibinfo {author} {\bibfnamefont {Eite}\ \bibnamefont {Tiesinga}},\
  }\bibfield  {title} {\enquote {\bibinfo {title} {Feshbach resonances in
  ultracold gases},}\ }\href {\doibase 10.1103/RevModPhys.82.1225} {\bibfield
  {journal} {\bibinfo  {journal} {Rev. Mod. Phys.}\ }\textbf {\bibinfo {volume}
  {82}},\ \bibinfo {pages} {1225--1286} (\bibinfo {year} {2010})}\BibitemShut
  {NoStop}%
\bibitem [{\citenamefont {Goldman}\ and\ \citenamefont
  {Dalibard}(2014)}]{goldman2014periodically}%
  \BibitemOpen
  \bibfield  {author} {\bibinfo {author} {\bibfnamefont {N.}~\bibnamefont
  {Goldman}}\ and\ \bibinfo {author} {\bibfnamefont {J.}~\bibnamefont
  {Dalibard}},\ }\bibfield  {title} {\enquote {\bibinfo {title} {Periodically
  driven quantum systems: Effective Hamiltonians and engineered gauge
  fields},}\ }\href {\doibase 10.1103/PhysRevX.4.031027} {\bibfield  {journal}
  {\bibinfo  {journal} {Phys. Rev. X}\ }\textbf {\bibinfo {volume} {4}},\
  \bibinfo {pages} {031027} (\bibinfo {year} {2014})}\BibitemShut {NoStop}%
\bibitem [{\citenamefont {Eckardt}\ and\ \citenamefont
  {Anisimovas}(2015)}]{Eckardt2015NJP}%
  \BibitemOpen
  \bibfield  {author} {\bibinfo {author} {\bibfnamefont {Andr{\'{e}}}\
  \bibnamefont {Eckardt}}\ and\ \bibinfo {author} {\bibfnamefont {Egidijus}\
  \bibnamefont {Anisimovas}},\ }\bibfield  {title} {\enquote {\bibinfo {title}
  {High-frequency approximation for periodically driven quantum systems from a
  Floquet-space perspective},}\ }\href {\doibase 10.1088/1367-2630/17/9/093039}
  {\bibfield  {journal} {\bibinfo  {journal} {New Journal of Physics}\ }\textbf
  {\bibinfo {volume} {17}},\ \bibinfo {pages} {093039} (\bibinfo {year}
  {2015})}\BibitemShut {NoStop}%
\bibitem [{\citenamefont {Rahav}\ \emph {et~al.}(2003)\citenamefont {Rahav},
  \citenamefont {Gilary},\ and\ \citenamefont {Fishman}}]{2003Effective}%
  \BibitemOpen
  \bibfield  {author} {\bibinfo {author} {\bibfnamefont {Saar}\ \bibnamefont
  {Rahav}}, \bibinfo {author} {\bibfnamefont {Ido}\ \bibnamefont {Gilary}}, \
  and\ \bibinfo {author} {\bibfnamefont {Shmuel}\ \bibnamefont {Fishman}},\
  }\bibfield  {title} {\enquote {\bibinfo {title} {Effective Hamiltonians for
  periodically driven systems},}\ }\href {\doibase 10.1103/PhysRevA.68.013820}
  {\bibfield  {journal} {\bibinfo  {journal} {Phys. Rev. A}\ }\textbf {\bibinfo
  {volume} {68}},\ \bibinfo {pages} {013820} (\bibinfo {year}
  {2003})}\BibitemShut {NoStop}%
\bibitem [{\citenamefont {Bukov}\ \emph {et~al.}(2015)\citenamefont {Bukov},
  \citenamefont {D'Alessio},\ and\ \citenamefont
  {Polkovnikov}}]{2015Universal}%
  \BibitemOpen
  \bibfield  {author} {\bibinfo {author} {\bibfnamefont {Marin}\ \bibnamefont
  {Bukov}}, \bibinfo {author} {\bibfnamefont {Luca}\ \bibnamefont {D'Alessio}},
  \ and\ \bibinfo {author} {\bibfnamefont {Anatoli}\ \bibnamefont
  {Polkovnikov}},\ }\bibfield  {title} {\enquote {\bibinfo {title} {Universal
  high-frequency behavior of periodically driven systems: from dynamical
  stabilization to Floquet engineering},}\ }\href {\doibase
  10.1080/00018732.2015.1055918} {\bibfield  {journal} {\bibinfo  {journal}
  {Advances in Physics}\ }\textbf {\bibinfo {volume} {64}},\ \bibinfo {pages}
  {139--226} (\bibinfo {year} {2015})}\BibitemShut {NoStop}%
\bibitem [{\citenamefont {Parameswaran}\ \emph {et~al.}(2013)\citenamefont
  {Parameswaran}, \citenamefont {Roy},\ and\ \citenamefont
  {Sondhi}}]{Fractional2013CRP}%
  \BibitemOpen
  \bibfield  {author} {\bibinfo {author} {\bibfnamefont {Siddharth~A.}\
  \bibnamefont {Parameswaran}}, \bibinfo {author} {\bibfnamefont {Rahul}\
  \bibnamefont {Roy}}, \ and\ \bibinfo {author} {\bibfnamefont {Shivaji~L.}\
  \bibnamefont {Sondhi}},\ }\bibfield  {title} {\enquote {\bibinfo {title}
  {Fractional quantum Hall physics in topological flat bands},}\ }\href
  {\doibase https://doi.org/10.1016/j.crhy.2013.04.003} {\bibfield  {journal}
  {\bibinfo  {journal} {Comptes Rendus Physique}\ }\textbf {\bibinfo {volume}
  {14}},\ \bibinfo {pages} {816--839} (\bibinfo {year} {2013})}\BibitemShut
  {NoStop}%
\bibitem [{\citenamefont {Marzari}\ \emph {et~al.}(2012)\citenamefont
  {Marzari}, \citenamefont {Mostofi}, \citenamefont {Yates}, \citenamefont
  {Souza},\ and\ \citenamefont {Vanderbilt}}]{2011Maximally}%
  \BibitemOpen
  \bibfield  {author} {\bibinfo {author} {\bibfnamefont {Nicola}\ \bibnamefont
  {Marzari}}, \bibinfo {author} {\bibfnamefont {Arash~A.}\ \bibnamefont
  {Mostofi}}, \bibinfo {author} {\bibfnamefont {Jonathan~R.}\ \bibnamefont
  {Yates}}, \bibinfo {author} {\bibfnamefont {Ivo}\ \bibnamefont {Souza}}, \
  and\ \bibinfo {author} {\bibfnamefont {David}\ \bibnamefont {Vanderbilt}},\
  }\bibfield  {title} {\enquote {\bibinfo {title} {Maximally localized Wannier
  functions: Theory and applications},}\ }\href {\doibase
  10.1103/RevModPhys.84.1419} {\bibfield  {journal} {\bibinfo  {journal} {Rev.
  Mod. Phys.}\ }\textbf {\bibinfo {volume} {84}},\ \bibinfo {pages}
  {1419--1475} (\bibinfo {year} {2012})}\BibitemShut {NoStop}%
\bibitem [{\citenamefont {Zhai}(2021)}]{zhai_2021}%
  \BibitemOpen
  \bibfield  {author} {\bibinfo {author} {\bibfnamefont {Hui}\ \bibnamefont
  {Zhai}},\ }\href {\doibase 10.1017/9781108595216} {\emph {\bibinfo {title}
  {Ultracold Atomic Physics}}}\ (\bibinfo  {publisher} {Cambridge University
  Press},\ \bibinfo {year} {2021})\BibitemShut {NoStop}%
\bibitem [{\citenamefont {Courteille}\ \emph {et~al.}(1998)\citenamefont
  {Courteille}, \citenamefont {Freeland}, \citenamefont {Heinzen},
  \citenamefont {van Abeelen},\ and\ \citenamefont
  {Verhaar}}]{Courteille1998PRL}%
  \BibitemOpen
  \bibfield  {author} {\bibinfo {author} {\bibfnamefont {Ph.}\ \bibnamefont
  {Courteille}}, \bibinfo {author} {\bibfnamefont {R.~S.}\ \bibnamefont
  {Freeland}}, \bibinfo {author} {\bibfnamefont {D.~J.}\ \bibnamefont
  {Heinzen}}, \bibinfo {author} {\bibfnamefont {F.~A.}\ \bibnamefont {van
  Abeelen}}, \ and\ \bibinfo {author} {\bibfnamefont {B.~J.}\ \bibnamefont
  {Verhaar}},\ }\bibfield  {title} {\enquote {\bibinfo {title} {Observation of
  a Feshbach resonance in cold atom scattering},}\ }\href {\doibase
  10.1103/PhysRevLett.81.69} {\bibfield  {journal} {\bibinfo  {journal} {Phys.
  Rev. Lett.}\ }\textbf {\bibinfo {volume} {81}},\ \bibinfo {pages} {69--72}
  (\bibinfo {year} {1998})}\BibitemShut {NoStop}%
\bibitem [{\citenamefont {Claussen}\ \emph {et~al.}(2003)\citenamefont
  {Claussen}, \citenamefont {Kokkelmans}, \citenamefont {Thompson},
  \citenamefont {Donley}, \citenamefont {Hodby},\ and\ \citenamefont
  {Wieman}}]{Claussen2003PRA}%
  \BibitemOpen
  \bibfield  {author} {\bibinfo {author} {\bibfnamefont {N.~R.}\ \bibnamefont
  {Claussen}}, \bibinfo {author} {\bibfnamefont {S.~J. J. M.~F.}\ \bibnamefont
  {Kokkelmans}}, \bibinfo {author} {\bibfnamefont {S.~T.}\ \bibnamefont
  {Thompson}}, \bibinfo {author} {\bibfnamefont {E.~A.}\ \bibnamefont
  {Donley}}, \bibinfo {author} {\bibfnamefont {E.}~\bibnamefont {Hodby}}, \
  and\ \bibinfo {author} {\bibfnamefont {C.~E.}\ \bibnamefont {Wieman}},\
  }\bibfield  {title} {\enquote {\bibinfo {title} {Very-high-precision
  bound-state spectroscopy near a ${}^{85}\mathrm{Rb}$ Feshbach resonance},}\
  }\href {\doibase 10.1103/PhysRevA.67.060701} {\bibfield  {journal} {\bibinfo
  {journal} {Phys. Rev. A}\ }\textbf {\bibinfo {volume} {67}},\ \bibinfo
  {pages} {060701(R)} (\bibinfo {year} {2003})}\BibitemShut {NoStop}%
\bibitem [{\citenamefont {Fukui}\ \emph {et~al.}(2005)\citenamefont {Fukui},
  \citenamefont {Hatsugai},\ and\ \citenamefont {Suzuki}}]{Fukui2005Chern}%
  \BibitemOpen
  \bibfield  {author} {\bibinfo {author} {\bibfnamefont {Takahiro}\
  \bibnamefont {Fukui}}, \bibinfo {author} {\bibfnamefont {Yasuhiro}\
  \bibnamefont {Hatsugai}}, \ and\ \bibinfo {author} {\bibfnamefont {Hiroshi}\
  \bibnamefont {Suzuki}},\ }\bibfield  {title} {\enquote {\bibinfo {title}
  {Chern numbers in discretized Brillouin zone: Efficient method of computing
  (spin) Hall conductances},}\ }\href {\doibase 10.1143/jpsj.74.1674}
  {\bibfield  {journal} {\bibinfo  {journal} {Journal of the Physical Society
  of Japan}\ }\textbf {\bibinfo {volume} {74}},\ \bibinfo {pages} {1674--1677}
  (\bibinfo {year} {2005})}\BibitemShut {NoStop}%
\bibitem [{\citenamefont {Viebahn}\ \emph {et~al.}(2021)\citenamefont
  {Viebahn}, \citenamefont {Minguzzi}, \citenamefont {Sandholzer},
  \citenamefont {Walter}, \citenamefont {Sajnani}, \citenamefont {G\"org},\
  and\ \citenamefont {Esslinger}}]{Suppressing2021Esslinger}%
  \BibitemOpen
  \bibfield  {author} {\bibinfo {author} {\bibfnamefont {Konrad}\ \bibnamefont
  {Viebahn}}, \bibinfo {author} {\bibfnamefont {Joaqu\'{\i}n}\ \bibnamefont
  {Minguzzi}}, \bibinfo {author} {\bibfnamefont {Kilian}\ \bibnamefont
  {Sandholzer}}, \bibinfo {author} {\bibfnamefont {Anne-Sophie}\ \bibnamefont
  {Walter}}, \bibinfo {author} {\bibfnamefont {Manish}\ \bibnamefont
  {Sajnani}}, \bibinfo {author} {\bibfnamefont {Frederik}\ \bibnamefont
  {G\"org}}, \ and\ \bibinfo {author} {\bibfnamefont {Tilman}\ \bibnamefont
  {Esslinger}},\ }\bibfield  {title} {\enquote {\bibinfo {title} {Suppressing
  dissipation in a Floquet-Hubbard system},}\ }\href {\doibase
  10.1103/PhysRevX.11.011057} {\bibfield  {journal} {\bibinfo  {journal} {Phys.
  Rev. X}\ }\textbf {\bibinfo {volume} {11}},\ \bibinfo {pages} {011057}
  (\bibinfo {year} {2021})}\BibitemShut {NoStop}%
\bibitem [{\citenamefont {Niu}\ \emph {et~al.}(1985)\citenamefont {Niu},
  \citenamefont {Thouless},\ and\ \citenamefont {Wu}}]{niu1985quantized}%
  \BibitemOpen
  \bibfield  {author} {\bibinfo {author} {\bibfnamefont {Qian}\ \bibnamefont
  {Niu}}, \bibinfo {author} {\bibfnamefont {D.~J.}\ \bibnamefont {Thouless}}, \
  and\ \bibinfo {author} {\bibfnamefont {Yong-Shi}\ \bibnamefont {Wu}},\
  }\bibfield  {title} {\enquote {\bibinfo {title} {Quantized Hall conductance
  as a topological invariant},}\ }\href {\doibase 10.1103/PhysRevB.31.3372}
  {\bibfield  {journal} {\bibinfo  {journal} {Phys. Rev. B}\ }\textbf {\bibinfo
  {volume} {31}},\ \bibinfo {pages} {3372--3377} (\bibinfo {year}
  {1985})}\BibitemShut {NoStop}%
\bibitem [{\citenamefont {Kj\"all}\ and\ \citenamefont
  {Moore}(2012)}]{PhysRevB.85.235137}%
  \BibitemOpen
  \bibfield  {author} {\bibinfo {author} {\bibfnamefont {Jonas~A.}\
  \bibnamefont {Kj\"all}}\ and\ \bibinfo {author} {\bibfnamefont {Joel~E.}\
  \bibnamefont {Moore}},\ }\bibfield  {title} {\enquote {\bibinfo {title} {Edge
  excitations of bosonic fractional quantum Hall phases in optical lattices},}\
  }\href {\doibase 10.1103/PhysRevB.85.235137} {\bibfield  {journal} {\bibinfo
  {journal} {Phys. Rev. B}\ }\textbf {\bibinfo {volume} {85}},\ \bibinfo
  {pages} {235137} (\bibinfo {year} {2012})}\BibitemShut {NoStop}%
\bibitem [{\citenamefont {Wen}(1990)}]{PhysRevB.41.12838}%
  \BibitemOpen
  \bibfield  {author} {\bibinfo {author} {\bibfnamefont {X.~G.}\ \bibnamefont
  {Wen}},\ }\bibfield  {title} {\enquote {\bibinfo {title} {Chiral Luttinger
  liquid and the edge excitations in the fractional quantum Hall states},}\
  }\href {\doibase 10.1103/PhysRevB.41.12838} {\bibfield  {journal} {\bibinfo
  {journal} {Phys. Rev. B}\ }\textbf {\bibinfo {volume} {41}},\ \bibinfo
  {pages} {12838--12844} (\bibinfo {year} {1990})}\BibitemShut {NoStop}%
\bibitem [{\citenamefont {Estienne}\ \emph {et~al.}(2015)\citenamefont
  {Estienne}, \citenamefont {Regnault},\ and\ \citenamefont
  {Bernevig}}]{PhysRevLett.114.186801}%
  \BibitemOpen
  \bibfield  {author} {\bibinfo {author} {\bibfnamefont {B.}~\bibnamefont
  {Estienne}}, \bibinfo {author} {\bibfnamefont {N.}~\bibnamefont {Regnault}},
  \ and\ \bibinfo {author} {\bibfnamefont {B.~A.}\ \bibnamefont {Bernevig}},\
  }\bibfield  {title} {\enquote {\bibinfo {title} {Correlation lengths and
  topological entanglement entropies of unitary and nonunitary fractional
  quantum Hall wave functions},}\ }\href {\doibase
  10.1103/PhysRevLett.114.186801} {\bibfield  {journal} {\bibinfo  {journal}
  {Phys. Rev. Lett.}\ }\textbf {\bibinfo {volume} {114}},\ \bibinfo {pages}
  {186801} (\bibinfo {year} {2015})}\BibitemShut {NoStop}%
\bibitem [{\citenamefont {Repellin}\ and\ \citenamefont
  {Goldman}(2019)}]{PhysRevLett.122.166801}%
  \BibitemOpen
  \bibfield  {author} {\bibinfo {author} {\bibfnamefont {C.}~\bibnamefont
  {Repellin}}\ and\ \bibinfo {author} {\bibfnamefont {N.}~\bibnamefont
  {Goldman}},\ }\bibfield  {title} {\enquote {\bibinfo {title} {Detecting
  fractional Chern insulators through circular dichroism},}\ }\href {\doibase
  10.1103/PhysRevLett.122.166801} {\bibfield  {journal} {\bibinfo  {journal}
  {Phys. Rev. Lett.}\ }\textbf {\bibinfo {volume} {122}},\ \bibinfo {pages}
  {166801} (\bibinfo {year} {2019})}\BibitemShut {NoStop}%
\bibitem [{\citenamefont {Price}\ and\ \citenamefont
  {Cooper}(2012)}]{PhysRevA.85.033620}%
  \BibitemOpen
  \bibfield  {author} {\bibinfo {author} {\bibfnamefont {H.~M.}\ \bibnamefont
  {Price}}\ and\ \bibinfo {author} {\bibfnamefont {N.~R.}\ \bibnamefont
  {Cooper}},\ }\bibfield  {title} {\enquote {\bibinfo {title} {Mapping the
  Berry curvature from semiclassical dynamics in optical lattices},}\ }\href
  {\doibase 10.1103/PhysRevA.85.033620} {\bibfield  {journal} {\bibinfo
  {journal} {Phys. Rev. A}\ }\textbf {\bibinfo {volume} {85}},\ \bibinfo
  {pages} {033620} (\bibinfo {year} {2012})}\BibitemShut {NoStop}%
\bibitem [{\citenamefont {Ra\ifmmode \check{c}\else
  \v{c}\fi{}i\ifmmode~\bar{u}\else \={u}\fi{}nas}\ \emph
  {et~al.}(2018)\citenamefont {Ra\ifmmode \check{c}\else
  \v{c}\fi{}i\ifmmode~\bar{u}\else \={u}\fi{}nas}, \citenamefont {\"Unal},
  \citenamefont {Anisimovas},\ and\ \citenamefont
  {Eckardt}}]{PhysRevA.98.063621}%
  \BibitemOpen
  \bibfield  {author} {\bibinfo {author} {\bibfnamefont {Mantas}\ \bibnamefont
  {Ra\ifmmode \check{c}\else \v{c}\fi{}i\ifmmode~\bar{u}\else \={u}\fi{}nas}},
  \bibinfo {author} {\bibfnamefont {F.~Nur}\ \bibnamefont {\"Unal}}, \bibinfo
  {author} {\bibfnamefont {Egidijus}\ \bibnamefont {Anisimovas}}, \ and\
  \bibinfo {author} {\bibfnamefont {Andr\'e}\ \bibnamefont {Eckardt}},\
  }\bibfield  {title} {\enquote {\bibinfo {title} {Creating, probing, and
  manipulating fractionally charged excitations of fractional Chern insulators
  in optical lattices},}\ }\href {\doibase 10.1103/PhysRevA.98.063621}
  {\bibfield  {journal} {\bibinfo  {journal} {Phys. Rev. A}\ }\textbf {\bibinfo
  {volume} {98}},\ \bibinfo {pages} {063621} (\bibinfo {year}
  {2018})}\BibitemShut {NoStop}%
\bibitem [{\citenamefont {Marzari}\ and\ \citenamefont
  {Vanderbilt}(1997)}]{Marzari1997PRB}%
  \BibitemOpen
  \bibfield  {author} {\bibinfo {author} {\bibfnamefont {Nicola}\ \bibnamefont
  {Marzari}}\ and\ \bibinfo {author} {\bibfnamefont {David}\ \bibnamefont
  {Vanderbilt}},\ }\bibfield  {title} {\enquote {\bibinfo {title} {Maximally
  localized generalized Wannier functions for composite energy bands},}\ }\href
  {\doibase 10.1103/PhysRevB.56.12847} {\bibfield  {journal} {\bibinfo
  {journal} {Phys. Rev. B}\ }\textbf {\bibinfo {volume} {56}},\ \bibinfo
  {pages} {12847--12865} (\bibinfo {year} {1997})}\BibitemShut {NoStop}%
\bibitem [{\citenamefont {Pizzi}\ \emph {et~al.}(2020)\citenamefont {Pizzi},
  \citenamefont {Vitale}, \citenamefont {Arita}, \citenamefont {Blügel},
  \citenamefont {Freimuth}, \citenamefont {G{\'{e}}ranton}, \citenamefont
  {Gibertini}, \citenamefont {Gresch}, \citenamefont {Johnson}, \citenamefont
  {Koretsune}, \citenamefont {Iba{\~{n}}ez-Azpiroz}, \citenamefont {Lee},
  \citenamefont {Lihm}, \citenamefont {Marchand}, \citenamefont {Marrazzo},
  \citenamefont {Mokrousov}, \citenamefont {Mustafa}, \citenamefont {Nohara},
  \citenamefont {Nomura}, \citenamefont {Paulatto}, \citenamefont
  {Ponc{\'{e}}}, \citenamefont {Ponweiser}, \citenamefont {Qiao}, \citenamefont
  {Thöle}, \citenamefont {Tsirkin}, \citenamefont {Wierzbowska}, \citenamefont
  {Marzari}, \citenamefont {Vanderbilt}, \citenamefont {Souza}, \citenamefont
  {Mostofi},\ and\ \citenamefont {Yates}}]{Pizzi2020}%
  \BibitemOpen
  \bibfield  {author} {\bibinfo {author} {\bibfnamefont {Giovanni}\
  \bibnamefont {Pizzi}}, \bibinfo {author} {\bibfnamefont {Valerio}\
  \bibnamefont {Vitale}}, \bibinfo {author} {\bibfnamefont {Ryotaro}\
  \bibnamefont {Arita}}, \bibinfo {author} {\bibfnamefont {Stefan}\
  \bibnamefont {Blügel}}, \bibinfo {author} {\bibfnamefont {Frank}\
  \bibnamefont {Freimuth}}, \bibinfo {author} {\bibfnamefont {Guillaume}\
  \bibnamefont {G{\'{e}}ranton}}, \bibinfo {author} {\bibfnamefont {Marco}\
  \bibnamefont {Gibertini}}, \bibinfo {author} {\bibfnamefont {Dominik}\
  \bibnamefont {Gresch}}, \bibinfo {author} {\bibfnamefont {Charles}\
  \bibnamefont {Johnson}}, \bibinfo {author} {\bibfnamefont {Takashi}\
  \bibnamefont {Koretsune}}, \bibinfo {author} {\bibfnamefont {Julen}\
  \bibnamefont {Iba{\~{n}}ez-Azpiroz}}, \bibinfo {author} {\bibfnamefont
  {Hyungjun}\ \bibnamefont {Lee}}, \bibinfo {author} {\bibfnamefont {Jae-Mo}\
  \bibnamefont {Lihm}}, \bibinfo {author} {\bibfnamefont {Daniel}\ \bibnamefont
  {Marchand}}, \bibinfo {author} {\bibfnamefont {Antimo}\ \bibnamefont
  {Marrazzo}}, \bibinfo {author} {\bibfnamefont {Yuriy}\ \bibnamefont
  {Mokrousov}}, \bibinfo {author} {\bibfnamefont {Jamal~I}\ \bibnamefont
  {Mustafa}}, \bibinfo {author} {\bibfnamefont {Yoshiro}\ \bibnamefont
  {Nohara}}, \bibinfo {author} {\bibfnamefont {Yusuke}\ \bibnamefont {Nomura}},
  \bibinfo {author} {\bibfnamefont {Lorenzo}\ \bibnamefont {Paulatto}},
  \bibinfo {author} {\bibfnamefont {Samuel}\ \bibnamefont {Ponc{\'{e}}}},
  \bibinfo {author} {\bibfnamefont {Thomas}\ \bibnamefont {Ponweiser}},
  \bibinfo {author} {\bibfnamefont {Junfeng}\ \bibnamefont {Qiao}}, \bibinfo
  {author} {\bibfnamefont {Florian}\ \bibnamefont {Thöle}}, \bibinfo {author}
  {\bibfnamefont {Stepan~S}\ \bibnamefont {Tsirkin}}, \bibinfo {author}
  {\bibfnamefont {Ma{\l}gorzata}\ \bibnamefont {Wierzbowska}}, \bibinfo
  {author} {\bibfnamefont {Nicola}\ \bibnamefont {Marzari}}, \bibinfo {author}
  {\bibfnamefont {David}\ \bibnamefont {Vanderbilt}}, \bibinfo {author}
  {\bibfnamefont {Ivo}\ \bibnamefont {Souza}}, \bibinfo {author} {\bibfnamefont
  {Arash~A}\ \bibnamefont {Mostofi}}, \ and\ \bibinfo {author} {\bibfnamefont
  {Jonathan~R}\ \bibnamefont {Yates}},\ }\bibfield  {title} {\enquote {\bibinfo
  {title} {Wannier90 as a community code: new features and applications},}\
  }\href {\doibase 10.1088/1361-648x/ab51ff} {\bibfield  {journal} {\bibinfo
  {journal} {Journal of Physics: Condensed Matter}\ }\textbf {\bibinfo {volume}
  {32}},\ \bibinfo {pages} {165902} (\bibinfo {year} {2020})}\BibitemShut
  {NoStop}%
\bibitem [{\citenamefont {Iba\~nez Azpiroz}\ \emph {et~al.}(2013)\citenamefont
  {Iba\~nez Azpiroz}, \citenamefont {Eiguren}, \citenamefont {Bergara},
  \citenamefont {Pettini},\ and\ \citenamefont {Modugno}}]{Tight2013PRA}%
  \BibitemOpen
  \bibfield  {author} {\bibinfo {author} {\bibfnamefont {Julen}\ \bibnamefont
  {Iba\~nez Azpiroz}}, \bibinfo {author} {\bibfnamefont {Asier}\ \bibnamefont
  {Eiguren}}, \bibinfo {author} {\bibfnamefont {Aitor}\ \bibnamefont
  {Bergara}}, \bibinfo {author} {\bibfnamefont {Giulio}\ \bibnamefont
  {Pettini}}, \ and\ \bibinfo {author} {\bibfnamefont {Michele}\ \bibnamefont
  {Modugno}},\ }\bibfield  {title} {\enquote {\bibinfo {title} {Tight-binding
  models for ultracold atoms in honeycomb optical lattices},}\ }\href {\doibase
  10.1103/PhysRevA.87.011602} {\bibfield  {journal} {\bibinfo  {journal} {Phys.
  Rev. A}\ }\textbf {\bibinfo {volume} {87}},\ \bibinfo {pages} {011602}
  (\bibinfo {year} {2013})}\BibitemShut {NoStop}%
\bibitem [{\citenamefont {{Str{\"a}ter}}\ and\ \citenamefont
  {{Eckardt}}(2016)}]{Interband2016ZNTA}%
  \BibitemOpen
  \bibfield  {author} {\bibinfo {author} {\bibfnamefont {Christoph}\
  \bibnamefont {{Str{\"a}ter}}}\ and\ \bibinfo {author} {\bibfnamefont
  {Andr{\'e}}\ \bibnamefont {{Eckardt}}},\ }\bibfield  {title} {\enquote
  {\bibinfo {title} {{Interband Heating Processes in a Periodically Driven
  Optical Lattice}},}\ }\href {\doibase 10.1515/zna-2016-0129} {\bibfield
  {journal} {\bibinfo  {journal} {Zeitschrift Naturforschung Teil A}\ }\textbf
  {\bibinfo {volume} {71}},\ \bibinfo {pages} {909--920} (\bibinfo {year}
  {2016})},\ \Eprint {http://arxiv.org/abs/1604.00850} {arXiv:1604.00850
  [cond-mat.quant-gas]} \BibitemShut {NoStop}%
\bibitem [{\citenamefont {Weinberg}\ \emph {et~al.}(2015)\citenamefont
  {Weinberg}, \citenamefont {\"Olschl\"ager}, \citenamefont {Str\"ater},
  \citenamefont {Prelle}, \citenamefont {Eckardt}, \citenamefont {Sengstock},\
  and\ \citenamefont {Simonet}}]{Multiphoton2015PRA}%
  \BibitemOpen
  \bibfield  {author} {\bibinfo {author} {\bibfnamefont {M.}~\bibnamefont
  {Weinberg}}, \bibinfo {author} {\bibfnamefont {C.}~\bibnamefont
  {\"Olschl\"ager}}, \bibinfo {author} {\bibfnamefont {C.}~\bibnamefont
  {Str\"ater}}, \bibinfo {author} {\bibfnamefont {S.}~\bibnamefont {Prelle}},
  \bibinfo {author} {\bibfnamefont {A.}~\bibnamefont {Eckardt}}, \bibinfo
  {author} {\bibfnamefont {K.}~\bibnamefont {Sengstock}}, \ and\ \bibinfo
  {author} {\bibfnamefont {J.}~\bibnamefont {Simonet}},\ }\bibfield  {title}
  {\enquote {\bibinfo {title} {Multiphoton interband excitations of quantum
  gases in driven optical lattices},}\ }\href {\doibase
  10.1103/PhysRevA.92.043621} {\bibfield  {journal} {\bibinfo  {journal} {Phys.
  Rev. A}\ }\textbf {\bibinfo {volume} {92}},\ \bibinfo {pages} {043621}
  (\bibinfo {year} {2015})}\BibitemShut {NoStop}%
\end{thebibliography}
\end{document}